\documentclass[12pt,reqno]{amsart}
\usepackage{amsmath,amssymb,amsfonts,amsthm}
\usepackage[mathscr]{eucal}
\usepackage[all]{xy}
\usepackage{hyperref}
\usepackage{setspace}
\usepackage{ytableau}
\allowdisplaybreaks

\author{V.A.~Abakumova, S.L.~Lyakhovich}

\address{Physics Faculty, Tomsk State University, Lenin ave. 36, Tomsk 634050, Russia}

\email{abakumova@phys.tsu.ru, \, sll@phys.tsu.ru}

\title{Dualisation of free fields}
\begin{document}
\maketitle

\begin{abstract}
We consider the general free field theory such that system of equations of motion includes a subsystem with a special property. If the subsystem is considered by itself, it would be a topological field theory having no local degrees of freedom. Various well-known field theories admit such a subsystem, including Einstein's gravity. As the subsystem is a topological theory, the general solution is a pure gauge, modulo global degrees of freedom. Gauge symmetry of the subsystem of linear theory can be explicitly found, providing the general solution to the subsystem. Substituting this solution into entire system, we arrive at an equivalent field theory where the role of the field variables is played by the gauge parameters of the topological subsystem. These fields can be viewed as the potentials for the original ones. A general procedure is proposed for constructing a ``parent action" which includes the fields of both dual formulations. At the level of this action, one can switch between dual formulations by imposing appropriate gauge fixing conditions. This general dualisation procedure is exemplified by massless and massive spin two fields. For the massless case, the hook type tensor serves as a potential for the metric, while for the massive case it is the fourth rank tensor with window Young diagram.
\end{abstract}

\section{Introduction}
In a broad sense, the dual field theories can be understood as the models whose mass shells are isomorphic modulo gauge symmetry. This covers transformations of the tensor fields by Hodge dualisation with respect to various subsets of indices in light-cone gauge with subsequent restoration of corresponding tensors in $d$ dimensions. Various cases are long known of the duality of this type, see, e.g. \cite{Curtright:1980yk}--\cite{Boulanger:2022arw} and references therein.  Another option is that dual fields can be connected by differential relations like strength tensors and their potentials, and these relations are not invertible at the local level even upon gauge fixing. All the dual formulations are equivalent at the classical level by definition, though their deformations, e.g. by inclusion of interaction into free theory,  are not necessarily equivalent.  The electromagnetic interaction of Dirac spinors, for example, being minimally switched on with the vector potential of electromagnetic field, cannot be locally reformulated in terms of field strength subject to Maxwell equations. Collections of massless spin one free fields do not admit Yang-Mills interactions if they are described by the strength tensors, not the potentials. Among the less obvious no-go results for interactions in dual formulations, we can refer to Curtright's dualisation of massless spin two \cite{Curtright:1980yk} which does not admit inclusion of interactions \cite{Bekaert:2002uh}.

In this article, we propose a general procedure of constructing dual formulations for the free field theories governed by quadratic Lagrangians. This procedure allows one to systematically introduce potentials for the fields proceeding from not quite restrictive assumptions about the structure of the free Lagrangian. For the simplest examples of massive and massless spin two, this already leads to previously unknown dualisations.

Let us briefly explain the key steps of the proposed dualisation procedure. 
We begin with the involutive closure of the equations of motion (EoMs). The closure implies that original Lagrangian equations are complemented by a certain set of their consequences. The system is considered involutive if it already includes all the lower order consequences of EoMs. The consequences can be added in involutive manner even if the original Lagrangian equations are already involutive. The involutive closure allows one to control the degree of freedom (DoF) number in manifestly covariant way \cite{Kaparulin:2012px}, so it provides a convenient starting point for inclusion of interactions.

As a second step, we assume that the involutive closure of original system includes a subsystem that would not describe any local DoF if considered on its own, independently of the entire system of field equations. We refer to these subsystems as topological theories as their dynamics are completely defined by global properties without any local DoF. This assumption is valid for various reasonable field theories\footnote{ For example, Einstein's system includes Nordstr\"om equation $R=\Lambda$ which is a topological theory in the above mentioned sense \cite{Abakumova:2022eoo}. Massive spin two equations for symmetric tensor admit the  first order differential consequences being the transversality equations $\partial_\mu h^{\mu\nu}=0$. These equations represent a topological theory.}. Proceeding from the assumption that the involutive closure of the EoMs includes a topological subsystem, we propose a systematic way of finding the complete gauge symmetry of the subsystem. This gauge symmetry is reducible, in general. At the free level, the gauge transformation is a general solution of the topological subsystem. Let us denote this pure gauge solution $\phi^i=\hat{\rho}{}^i{}_A\omega^A$, where $\phi^i$ are original fields, $\hat{\rho}{}^i{}_A$ are gauge generators, being local differential operators, and $\omega^A$ are gauge parameters, cf. (\ref{GenSolTau}). Substituting the general solution of the subsystem into the entire system of field equations, one doesn't miss or introduce any information about dynamics. This ``partially solved'' system of field equations is equivalent to the original one by construction. The unknown fields of this equivalent system are the gauge parameters $\omega^A$ of the topological subsystem. These new fields $\omega^A$ can be viewed as potentials for the original fields $\phi^i$, while equations for the potentials can be considered as dual to the original system of EoMs. The dual equations do not necessarily admit variational principle even though the original equations are Lagrangian. These first two steps are presented in the next section.

As a third step, we propose  a procedure of embedding of the original EoMs and their dual counterpart into a uniform ``parent theory" which is Lagrangian by construction.  The ``parent action" involves both original fields and their potentials\footnote{For discussion of the role of ``parent actions" for dual field theories, see \cite{Boulanger:2003vs}.}. This action has a more wide gauge symmetry than both dual formulations. The embedding follows the general method of inclusion of Stueckelberg fields \cite{Lyakhovich:2021lzy}, \cite{Abakumova:2021evc} which proceeds from the involutive closure of the original Lagrangian system. The procedure implies to introduce the Stueckelberg field for every consequence of Lagrangian equations included into the involutive closure. As a starting point for the embedding, we choose a certain extension of the original involutive closure. This extension includes, besides the topological subsystem, some extra consequences, being gauge variations of the original action w.r.t. gauge transformations of the topological subsystem.  From the perspective of this embedding, the potentials $\omega^A$ for the original fields $\phi^i$ appear as the Stueckelberg fields associated to these extra consequences. This procedure is explained in Section \ref{Sec:3}. Gauge symmetry of the ``parent theory" can be fixed in different ways. In particular, one can impose a gauge condition which kills all the Stueckelberg fields reproducing the original theory. There is an alternative set of gauge conditions which kills the original fields reproducing the dual theory in terms of potentials $\omega^A$.

The general construction of dualisation is exemplified by specific models in Section \ref{Sec:4}. As a warm-up, we first consider the dualisation of Proca's Lagrangian for massive spin one. The parent Lagrangian for this dualisation in $d=4$ has been already reported in the article \cite{Abakumova:2021evc}, though without relation to the general scheme proposed in the present article. We further consider dualisation of massive and massless spin two theories proceeding from Lagrangians for symmetric second rank tensors. As a result, we arrive to the parent theory where the massless spin two is described by the symmetric tensor and its potential being third rank tensor with hook Young diagram. For the massive spin two, the gauge potential is the fourth rank tensor with window Young diagram.

The proposed dualisation procedure may seem restricted to the free level. In conclusion, we discuss the perspectives of extending the construction beyond linear theory.

\section{Gauge symmetry of topological subsystem and dualisation}\label{Sec:2}

Given set of fields $\phi^i(x)$, let us consider the most general translation-invariant quadratic action
\begin{equation}\label{S}
\displaystyle S[\phi]=\int d^d x\,\mathcal{L}\,, \quad
\mathcal{L}=\frac{1}{2}\phi^i \hat{M}_{ij}\phi^j\,,
\end{equation}
where the matrix elements $\hat{M}_{ij}$ are differential operators, being polynomials of partial derivatives\footnote{In this paper, the hat symbol means that corresponding quantity is a differential operator with constant coefficients.}
\begin{equation}\label{M}
\displaystyle \hat{M}_{ij}(\partial)=\sum_{k=0}^{n}M_{ij}^{\mu_1\dots\mu_k}\partial_{\mu_1}\ldots\partial_{\mu_k}\, , \quad M_{ij}^{\mu_1\dots\mu_k}=const\, , \quad \partial_\mu=\frac{\partial}{\partial x^\mu}\,.
\end{equation}
Without restriction of generality, the matrix $\hat{M}_{ij}$ can be assumed Hermitian in the sense that it remains unchanged under transposition complemented with reflection of the partial derivatives:
\begin{equation}\label{Hermitian}
\displaystyle \hat{M}_{ij}(\partial)=\hat{M}_{ji}(-\partial) \equiv \hat{M}^\dagger_{ji}(\partial) \,.
\end{equation}
Any anti-Hermitian part would add a total divergence to Lagrangian (\ref{S}).
Given the action, EoMs read
\begin{equation}\label{EoMs}
\displaystyle E_i\equiv\frac{\delta S}{\delta \phi^i}=\hat{M}_{ij}\phi^j=0\,.
\end{equation}
Let us briefly remind the general algebraic setup for the gauge symmetry of the quadratic action. Consider the kernel of the matrix $\hat{M}$,
\begin{equation}\label{kerM}
\displaystyle\hat{U}{}^j \in\ker\hat{M}_{ij}\quad\Leftrightarrow\quad \hat{M}_{ij}\hat U{}^j=0 \, .
\end{equation}
Let $\hat{R}^i{}_\alpha$, $\alpha=1,\dots, r$, be the generating set in the kernel of $\hat{M}_{ij}$, i.e.
\begin{equation}\label{R}
\displaystyle \hat{U}^i\in\ker\hat{M}_{ij} \quad \Leftrightarrow \quad\exists\,\hat{U}^\alpha: \quad \hat{U}^i=\hat{R}^i{}_\alpha\hat{U}^\alpha \, .
\end{equation}
Given the gauge generators, the action (\ref{S}) is invariant under the following gauge transformations:
\begin{equation}\label{gtr-R}
\displaystyle \delta_\epsilon \phi^i=\hat{R}^i{}_\alpha\epsilon^\alpha(x)\,, \quad \delta_\epsilon S[\phi]\equiv 0\,, \quad \forall\, \epsilon^\alpha(x)\, .
\end{equation}
The generating set (\ref{R}) can be overcomplete, then the gauge symmetry is reducible. From the perspective of commutative algebra, the issue of finding all the gauge generators and the sequence of symmetry for symmetry transformations reduces to the syzygy problem, see in Section 4 of the article \cite{Francia:2013sca}. The syzygy problem can be always solved by various methods \cite{Eisenbud:2005}.  Also mention that for a wide class of tensorial equations, the generalised Poincar\'e lemma of the article \cite{Bekaert:2002dt} can provide useful tools for finding the sequence of reducible gauge transformations. So one can assume that complete reducible gauge symmetry is known. Given the gauge symmetry generators and the hermiticity of $\hat{M}$ (\ref{Hermitian}), the Noether identities for Lagrangian equations (\ref{EoMs}) are generated by the conjugate operators for $\hat{R}$'s:
\begin{equation}\label{NI's}
\displaystyle  \hat{R}^{\dagger}_\alpha{}^iE_i\equiv 0\,.
\end{equation}

Let us consider a set of differential  consequences of original EoM (\ref{EoMs})
\begin{equation}\label{diffcons}
\displaystyle \tau_a\equiv\hat{\Gamma}^\dagger_a{}^iE_i=\hat{\Gamma}^\dagger_a{}^i\hat{M}_{ij}\phi^j\approx 0\,,
\end{equation}
where $\approx$ means on-shell equality.
The system
\begin{equation}\label{Inv}
\displaystyle E_i=0\,, \quad \tau_a=0\,,
\end{equation}
is equivalent to (\ref{EoMs}) as the mass shell remains the same. We assume that (\ref{Inv}) is  involutively closed, i.e. it contains all the lower order consequences of (\ref{EoMs}). Besides the Noether identities (\ref{NI's}), the involutive closure (\ref{Inv}) by construction enjoys the extra gauge identities
\begin{equation}\label{GI-closure}
\displaystyle \hat{\Gamma}^\dagger_a{}^i{}E_i-\tau_a \equiv 0\,.
\end{equation}
These identities are unrelated to any gauge symmetry. This does not contradict the second Noether theorem as the involutive closure (\ref{Inv}) is a non-Lagrangian system. Much like gauge symmetries, the gauge identities can be reducible, and the identities of identities can have further reducibility.

Given the gauge algebra of involutive system, including identities, symmetries, and their reducibilities, one can count the DoF number in a manifestly covariant manner \cite{Kaparulin:2012px}, while the original Lagrangian equations do not admit the direct DoF count. Furthermore, it is the gauge algebra of the involutive closure, not of the original Lagrangian system, which controls consistent inclusion of interactions \cite{Kaparulin:2012px}. If the original Lagrangian system is involutive, and the gauge symmetry is irreducible, manifestly covariant DoF count can be made in a more simple way \cite{Henneaux:1990au}, and Lagrangian setup is sufficient for consistent inclusion of interactions \cite{Barnich:1993vg}, \cite{Henneaux:1997bm}.

There is a systematic way to cast the involutive closure of the Lagrangian system back into Lagrangian framework by inclusion of Stueckelberg fields \cite{Lyakhovich:2021lzy}.  The method of the article \cite{Lyakhovich:2021lzy} can be viewed as manifestly covariant counterpart of the procedure of converting the Hamiltonian second class constraints into first class ones by introducing ``conversion variable'' for every second class constraint \cite{Faddeev:1986pc}--\cite{Batalin:2005df}. For the general covariant Stueckelberg embedding method, the Stueckelberg field, being the Lagrangian analogue of Hamiltonian ``conversion variable", is assigned to every consequence included into involutive closure of Lagrangian system. If the consequences $\tau_a$ (\ref{diffcons}) are reducible, the method is adjusted by certain modification that results in reducible Stueckelberg gauge symmetry \cite{Abakumova:2021evc}. In this article we propose a general scheme for constructing dual formulations of given free field theories that proceeds from involutive closure of Lagrangian equations. Then, we cast the dual formulation back into Lagrangian setup by introducing Stueckelberg fields along the lines of the method proposed in \cite{Abakumova:2021evc}.

As a first step to constructing dual formulation for the original free field theory, let us assume that the set of consequences (\ref{diffcons}), being considered by itself, irrespectively to entire system (\ref{Inv}),  admit a more wide gauge symmetry than the original system:
\begin{equation}\label{gtr-0}
\displaystyle \delta^{(0)}_\zeta \phi^i=\hat{\rho}{}^i{}_A\zeta^A\,, \quad
\delta^{(0)}_\zeta \tau_a \equiv 0\,, \,\, \forall\, \zeta^A \,.
\end{equation}
This symmetry is equivalent to the following relation:
\begin{equation}\label{gtr-00}
 \hat{\Gamma}^\dagger_a{}^i\hat{M}_{ij}\hat{\rho}{}^i{}_A\equiv0\,,
\end{equation}
where operators $\hat{\Gamma}^\dagger_a{}^i$ are introduced  by relations (\ref{diffcons}) as generators of differential consequences.

Note that original  gauge transformation (\ref{gtr-R}) is included in (\ref{gtr-0}), i.e.
\begin{equation}
\displaystyle \exists\,\, \hat{R}^A{}_\alpha:\,\,\hat{R}^i{}_\alpha= \hat{R}^A{}_\alpha \, \hat{\rho}^i{}_A\,.
\end{equation}
The gauge transformations (\ref{gtr-0}) can be further reducible. This means, the set of generators $\hat{\rho}^i{}_A$ of gauge symmetry transformations (\ref{gtr-0}) of the system of consequences (\ref{diffcons}) can be overcomplete, and they admit identities. The generators of these identities can be further reducible, etc. The reducibility leads to gauge symmetry of the gauge parameters, i.e. one can transform the gauge parameters without affecting the transformation (\ref{gtr-0}) of the original fields. The most general sequence of symmetry for symmetry gauge transformations reads
\begin{eqnarray}\label{gtr-1}
\displaystyle \delta^{(1)}_{\zeta_1}\zeta^A=\hat{\rho}^A{}_{A_1}\zeta_1^{A_1}\,: \quad   \delta^{(1)}_{\zeta_1}\delta^{(0)}_\zeta \phi^i\equiv 0\,,\,\, \forall\,\zeta_1^{A_1} \,\, \Leftrightarrow \,\, \hat{\rho}^i{}_A\hat{\rho}^A{}_{A_1}\equiv0\,;\\[3mm]\label{gtr-2}
\displaystyle \delta^{(2)}_{\zeta_2}\zeta_1^{A_1}=\hat{\rho}^{A_1}{}_{A_2}\zeta_2^{A_2}\,: \quad \delta^{(2)}_{\zeta_2}\delta^{(1)}_{\zeta_1}\zeta^A=0\,, \,\, \forall\,\zeta_2^{A_2} \,\, \Leftrightarrow \,\, \hat{\rho}^A{}_{A_1}\hat{\rho}^{A_1}{}_{A_2}\equiv0\,;\\[3mm]
\label{gtr-k}
\displaystyle \delta^{(k)}_{\zeta_k}\zeta_{k-1}^{A_{k-1}}=\hat{\rho}^{A_{k-1}}{}_{A_k}\zeta_k^{A_k}\,: \quad \delta^{(k)}_{\zeta_k}\delta^{(k-1)}_{\zeta_{k-1}}\zeta_{k-2}^{A_{k-2}}=0\,,\,\, \forall\,\zeta_k^{A_k} \,\, \Leftrightarrow \,\, \hat{\rho}^{A_{k-2}}{}_{A_{k-1}}\hat{\rho}^{A_{k-1}}{}_{A_k}\equiv 0\,,
\end{eqnarray}
where $k=3,\ldots, n$, and the reducibility order $n$ depends on the model and the dimension of the space-time. To comply with Hilbert's syzygy theorem \cite{Eisenbud:2005}, the reducibility order is bounded by the space-time dimension: $n\leq d$ . Gauge parameters of $k$-th stage $\zeta_k^{A_k}$, $k=1,\ldots,n$,  are assumed to be arbitrary functions of $x$.  Once the fields are tensors, and the involutive closure of EoMs (\ref{Inv}) is a covariant system, the gauge parameters have to be tensors.

Let us further assume that consequences $\tau_{a}$ (\ref{diffcons}), being considered irrespectively to the complete system (\ref{Inv}), form a topological subsystem of EoMs in the sense that it does not have any local DoF.  Given the complete sequences of gauge symmetries and identities of the involutive system, one can immediately count the DoF \cite{Kaparulin:2012px}, so this assumption can be always verified.

The gauge symmetry generators $\hat{\rho}^i{}_A$ (\ref{gtr-0}) of the topological subsystem do not result in extra gauge identities (besides the original Noether identities (\ref{NI's})) between the original EoM's, hence they define more non-trivial higher order consequences:
\begin{equation}\label{TA}
\displaystyle \mathcal{T}_A\equiv\hat{\rho}^\dagger_A{}^iE_i=\hat{\rho}^\dagger_A{}^i\hat{M}_{ij}\phi^j\approx 0\,,
\end{equation}
where $\hat{\rho}^\dagger_A{}^i$ is a conjugate to $\hat{\rho}^i{}_A$ in (\ref{gtr-0}). By construction, there exist gauge identities between the above consequences and original Lagrangian equations,
\begin{equation}\label{GI-TA}
\displaystyle \mathcal{T}_A-\hat{\rho}^\dagger_A{}^iE_i\equiv0\,.
\end{equation}
The set of generators $\hat{\rho}^\dagger_A{}^i$ of consequences (\ref{TA}) is overcomplete,
\begin{equation}\label{TA-1-1}
\displaystyle \hat{\rho}^\dagger_{A_1}{}^A\hat{\rho}^\dagger_A{}^i\equiv0\,,
\end{equation}
where $\hat{\rho}^\dagger_{A_1}{}^A$ is a conjugate to $\hat{\rho}{}^A{}_{A_1}$ in (\ref{gtr-1}). This leads to identities between consequences (\ref{TA}),
\begin{equation}\label{TA-1-2}
\displaystyle \hat{\rho}^\dagger_{A_1}{}^A\mathcal{T}_A\equiv0\,.
\end{equation}
These identities, because of (\ref{gtr-2})--(\ref{gtr-k}), admit further reducibility,
\begin{eqnarray}\label{TA-2}
\displaystyle \hat{\rho}^\dagger_{A_2}{}^{A_1}\hat{\rho}^\dagger_{A_1}{}^A\equiv0\,;
\\[3mm]\label{TA-k}
\displaystyle \hat{\rho}^\dagger_{A_k}{}^{A_{k-1}}\hat{\rho}^\dagger_{A_{k-1}}{}^{A_{k-2}}\equiv0\,,
\end{eqnarray}
where $k=3,\ldots,n$, and $\hat{\rho}^\dagger_{A_2}{}^{A_1}$ and $\hat{\rho}^\dagger_{A_k}{}^{A_{k-1}}$ are conjugate to $\hat{\rho}^{A_1}{}_{A_2}$ and $\hat{\rho}^{A_{k-1}}{}_{A_k}$, cf. (\ref{gtr-2})--(\ref{gtr-k}).

The system $E_i=0$ (\ref{EoMs}), being completed by their consequences,
\begin{equation}\label{Inv1}
E_i=0\,, \quad \tau_a=0\,, \quad \mathcal{T}_A=0\,,
\end{equation}
is obviously equivalent to original Lagrangian equations. Once involutive closure (\ref{Inv}) is complemented in (\ref{Inv1}) by further consequences (\ref{TA}) constructed in the special way described above, this does not break involutivity. This fact will be proven elsewhere, here it is a conjecture which is confirmed by examples. The involutive system (\ref{Inv1}) is non-Lagrangian. Involutive closure of any Lagrangian system can be systematically brought back to Lagrangian framework without spoiling involutivity by inclusion of Stueckelberg fields, see in \cite{Lyakhovich:2021lzy}, \cite{Abakumova:2021evc}. The Stueckelberg procedure for (\ref{Inv1}) is implemented in the next section. To finalise this section, we explain how the system (\ref{Inv1}) leads to dualisation of the theory at non-Lagrangian level.

The equations $\tau_a=0$, being considered irrespectively to the entire system (\ref{Inv1}), do not carry any local DoF, hence the general solution to these equations is a pure gauge,
\begin{equation}\label{GenSolTau}
  \phi^i=\hat{\rho}^i{}_A\omega^A(x)\,, \,\,\forall\,\omega^A \,,
\end{equation}
cf. (\ref{gtr-0}), where the arbitrary gauge parameters $\zeta^A$ are considered as new independent fields, denoted $\omega^A$. Substituting the general solution for the subsystem $\tau_a=0$ into the rest of equations (\ref{Inv}), we arrive to the equations for the new fields $\omega^A$,
\begin{equation}\label{dual}
\overline{E}_i\equiv \hat{M}_{ij}\hat{\rho}^j{}_A\omega^A\approx 0\, .
\end{equation}
This system can be understood as the adjoint to the equations $\mathcal{T}_A=0$ (\ref{TA}) in the sense that operator of equations (\ref{dual}) is related to that of (\ref{TA}) by conjugation. From another perspective, equations (\ref{dual}) for the fields $\omega^A$ can be considered as dual for the Lagrangian equations (\ref{EoMs}) for the fields $\phi^i$ because the mass shells of these systems are by construction isomorphic, modulo gauge symmetry. The fields $\omega^A$ obeying the higher order equations (\ref{dual}) can be viewed as ``potentials'' for the fields $\phi^i$ subject to the lower order equations (\ref{EoMs}). The fields and their ``potentials'' are connected by relations (\ref{GenSolTau}).

Notice that gauge symmetry transformations of dual system (\ref{dual}) coincide with (\ref{gtr-1})--(\ref{gtr-k}), i.e.
\begin{eqnarray}\label{gst-dual-1}
\displaystyle \delta^{(0)}_{\xi_1}\omega^A=\hat{\rho}^A{}_{A_1}\xi_1^{A_1}\,;
\\\label{gst-dual-k}
\displaystyle \delta^{(k-1)}_{\xi_k}\xi_{k-1}^{A_{k-1}}=\hat{\rho}^{A_{k-1}}{}_{A_k}\xi_k^{A_k}\,,
\end{eqnarray}
where $k=2,\ldots, n$, and the gauge parameters $\xi$ are arbitrary tensors with certain symmetry type.

The gauge identities of the dual equations (\ref{dual}) are inherited from the Noether identities of original equations (\ref{NI's}), and from the fact that relations (\ref{GenSolTau}) solve the consequences $\tau_a=0$ of the EoMs (\ref{diffcons}):
\begin{equation}\label{GI-Dual1}
\displaystyle \hat{R}^\dagger_\alpha{}^i\overline{E}_i\equiv\hat{R}^\dagger_\alpha{}^i\hat{M}_{ij}\hat{\rho}^j{}_A\omega^A\equiv0\,;
\end{equation}
\begin{equation}\label{GI-Dual2}
\displaystyle \hat{\Gamma}^\dagger_a{}^i\overline{E}_i\equiv \hat{\Gamma}^\dagger_a{}^i\hat{M}_{ij}\hat{\rho}^j{}_A\omega^A\equiv\delta^{(0)}_\omega(\tau_a)\equiv0\,.
\end{equation}
As one can see, if the linear field equations admit a subsystem such that does not have any DoF being considered in itself, the original Lagrangian theory (\ref{EoMs}) of the fields $\phi^i$ with gauge symmetries (\ref{gtr-R}) is equivalently reformulated in terms of their ``potentials''  $\omega^A$ (\ref{GenSolTau}) with EoMs (\ref{dual}), gauge symmetries (\ref{gst-dual-1})--(\ref{gst-dual-k}) and gauge identities (\ref{GI-Dual1}), (\ref{GI-Dual2}). There is no pairing between gauge symmetry transformations and gauge identities  because the equations (\ref{dual}) are non-Lagrangian.

This general procedure of dualisation is reported here for the first time, though the two specific simple examples have been previously noted in \cite{Abakumova:2021evc}, \cite{Abakumova:2022eoo}. The simplest example is provided by Proca equations for massive spin one. The equations admit the consequence $\tau=m^2\partial_\mu A^\mu\approx 0$ (cf. (\ref{diffcons})), that allows one to reformulate the theory in terms of ``potential'' being antisymmetric tensor $A^\mu=\partial_\nu B^{\mu\nu}, \, B^{\mu\nu}=-\,B^{\nu\mu}$, cf. (\ref{GenSolTau}). This example is detailed in the beginning of Section \ref{Sec:4}. Less obvious example of the dualisation of this type is provided by linearised Einstein's equations. Their trace is the linearised Nordstr\"om equation, which is a topological theory. The gauge symmetry of Nordstr\"om equation is parameterised by the third rank tensor with hook Young diagram. This tensor can be viewed as a potential for linearised Einstein's metric, cf. (\ref{GenSolTau}). Resolving linearised Nordstr\"om equation in terms of the hook, one arrives at the massless spin two theory in terms of the hook tensor field \cite{Abakumova:2022eoo}. In Section \ref{Sec:4}, we elaborate on the examples.

In the next section we consider the Stueckelberg procedure which allows one to cast the non-Lagrangian dual formulation (\ref{dual}) back to the Lagrangian setup.

\section{Construction of Stueckelberg Lagrangian for dual formulation}\label{Sec:3}

The general procedure of inclusion of Stueckelberg fields proceeds from involutive closure of the original Lagrangian system \cite{Lyakhovich:2021lzy}, \cite{Abakumova:2021evc}. Every consequence of Lagrangian equations included into the closure is assigned with the Stueckelberg field. The Stueckelberg action is sought for as the power series in Stueckelberg fields. The initial term of the expansion is the original action, while the first order is the sum of products of the Stueckelberg fields to corresponding consequences. The Stueckelberg gauge generators begin with the generators of extra gauge identities (see (\ref{GI-closure})) between the Lagrangian equations and their consequences included into the involutive closure \cite{Lyakhovich:2021lzy}. The procedure does not have obstructions \cite{Lyakhovich:2021lzy}, and it allows one to iteratively construct both the action and its gauge symmetry. For the free field theories, the iterative procedure have to terminate at quadratic terms in Stueckelberg fields in the action, while the gauge generators are field-independent, so they are immediately defined by zero-order boundary conditions.

Given the involutive closure of Lagrangian equations (\ref{Inv1}), we introduce Stueckelberg fields $\sigma^a$ and $\omega^A$ for the consequences $\tau_a$ and $\mathcal{T}_A$, respectively, and following the procedure proposed in \cite{Abakumova:2021evc}, we seek for the Stueckelberg action of the form
\begin{equation}\label{St-action-1}
\displaystyle \mathcal{S}[\phi, \sigma,\omega]=S[\phi]+\sigma^a\tau_a+\omega^A\mathcal{T}_A+\ldots\,.
\end{equation}
Given the gauge identities (\ref{GI-closure}), (\ref{GI-TA}) and (\ref{TA-1-2}), the general recipe of \cite{Abakumova:2021evc} leads to the Stueckelberg gauge transformations
\begin{equation}\label{StGS}
\displaystyle \delta^{(0)}\phi^i=-\,\hat{\Gamma}^i{}_a\mathcal{E}^a-\hat{\rho}^i{}_A\mathcal{E}^A\,, \quad \delta^{(0)}\sigma^a=\mathcal{E}^a\,, \quad \delta^{(0)}\omega^A=\mathcal{E}^A-\hat{\rho}^A{}_{A_1}\xi^{A_1}\,,
\end{equation}
where $\mathcal{E}^a, \mathcal{E}^A, \xi^{A_1} $ are the gauge parameters, being arbitrary functions of $x$. Notice that there is no pairing between gauge parameters and consequences included into the involutive closure (\ref{Inv1}) because of the identities between the consequences (\ref{TA-1-1}). This is a distinction from the case with irreducible consequences considered in \cite{Lyakhovich:2021lzy}.

Gauge transformations (\ref{StGS}) are reducible,
\begin{eqnarray}\label{StGS-reduce-1}
\displaystyle \delta^{(1)}\mathcal{E}^A=\hat{\rho}^A{}_{A_1}\mathcal{E}^{A_1}\,, \quad \delta^{(1)}\mathcal{E}^a=0\,, \quad \delta^{(1)}\xi^{A_1}=\mathcal{E}^{A_1}-\hat{\rho}^{A_1}{}_{A_2}\xi^{A_2}\,;\\\label{StGS-reduce-2}
\displaystyle \delta^{(k)}\mathcal{E}^{A_{k-1}}=\hat{\rho}^{A_{k-1}}{}_{A_{k}}\mathcal{E}^{A_k}\,, \quad \delta^{(k)}\xi^{A_{k}}=\mathcal{E}^{A_k}-\hat{\rho}^{A_k}{}_{A_{k+1}}\xi^{A_{k+1}}\,,
\quad k=2,\ldots, n-1\,;\\\label{StGS-reduce-k}
\displaystyle \delta^{(n)}\mathcal{E}^{A_{n-1}}=\hat{\rho}^{A_{n-1}}{}_{A_{n}}\mathcal{E}^{A_{n}}\,, \quad \delta^{(n)}\xi^{n}=\mathcal{E}^{A_{n}}\,,
\end{eqnarray}
with $\mathcal{E}^{A_1}$, $\mathcal{E}^{A_k}$ and $\xi^{A_{k}}$, $k=2,\ldots, n$, being gauge parameters of symmetry for symmetry transformations, where $n$ depends on the model and the dimension of the space-time.

Given the gauge transformations above, and the first order of the action w.r.t. to the Stueckelberg fields (\ref{St-action-1}), the requirement of gauge invariance defines the second order in $\sigma^a,\omega^A$. As a result, the Stueckelberg action reads
\begin{equation}\label{SSt}
\displaystyle \mathcal{S}[\phi,\sigma,\omega]=\int dx\,\mathfrak{L}\,, \quad \mathfrak{L}=\frac{1}{2}\big(\phi^i+\hat{\Gamma}^i{}_a\sigma^a+\hat{\rho}^i{}_A\omega^A\big)\hat{M}_{ij}\big(\phi^j+\hat{\Gamma}^j{}_b\sigma^b+\hat{\rho}^j{}_B\omega^B\big)\,,
\end{equation}
where $\phi^i$ are the original fields, while $\sigma^a$ and $\omega^A$ play the role of Stueckelberg fields. The corresponding equations of motion read
\begin{eqnarray}\label{EoMStphi}
\displaystyle \frac{\delta \mathcal{S}}{\delta \phi^i}\equiv\hat{M}_{ij}\big(\phi^j+\hat{\Gamma}^j{}_b\sigma^b+\hat{\rho}^j{}_B\omega^B\big)=0\,;\\\label{EoMStsigma}
\displaystyle \frac{\delta \mathcal{S}}{\delta \sigma^a}\equiv\hat{\Gamma}^\dagger_a{}^i\hat{M}_{ij}\big(\phi^j+\hat{\Gamma}^j{}_b\sigma^b+\hat{\rho}^j{}_B\omega^B\big)=0\,;\\\label{EoMStomega}
\displaystyle \frac{\delta \mathcal{S}}{\delta \omega^A}\equiv\hat{\rho}^\dagger_A{}^i\hat{M}_{ij}\big(\phi^j+\hat{\Gamma}^j{}_b\sigma^b+\hat{\rho}^j{}_B\omega^B\big)=0\,.
\end{eqnarray}

The reducible Stueckelberg gauge symmetry (\ref{StGS}) admits gauge-fixing conditions
\begin{equation}\label{gauge1}
\displaystyle \sigma^a=0\,, \quad \omega^A=0\,.
\end{equation}
These conditions eliminate all the Stueckelberg fields, and equations  (\ref{EoMStphi}), (\ref{EoMStsigma}), (\ref{EoMStomega}) reproduce the involutive closure (\ref{Inv1}) of the original equations. One can see that the gauge orbit is transverse to the surface of this gauge, so it is admissible indeed.

Let us note that Stueckelberg gauge symmetry (\ref{StGS}) can be fixed by another set of conditions
\begin{equation}\label{gauge2}
\displaystyle \phi^i=0\,, \quad \sigma^a=0\,, \quad \hat{\rho}^\dagger{}^{A_1}{}_A\omega^A=0\,,
\end{equation}
where $\hat{\rho}^\dagger{}^{A_1}{}_A$ is $\hat{\rho}^\dagger{}_{A_1}{}^A$ with indices raised and lowered by the space-time metric. This gauge kills the original fields $\phi^i$, and the Stueckelberg fields $\sigma^a$ related to the topological subsystem. The Stueckelberg fields $\omega^A$ related to consequences $\mathcal{T}_A$ (\ref{TA}) included into the involutive closure of EoMs (\ref{Inv1}) remain dynamical and obey the dual equations (\ref{dual}), while their gauge symmetry is fixed by the last set of gauge conditions in (\ref{gauge2}).

From the perspective of Stueckelberg embedding of the involutive closure (\ref{Inv1})  of original system, the potentials $\omega^A$, being the fields of dual formulation (\ref{dual}), serve as the Stueckelberg fields introduced to gauge the consequences $\mathcal{T}_A$ (\ref{TA}). The Stueckelberg formulation (\ref{SSt}) can be considered as a ``parent theory'' which encompasses both the Lagrangian system (\ref{S}) in terms of fields $\phi^i$ and non-Lagrangian dual system of equations (\ref{dual}) for the potentials $\omega^A$. This ``parent theory" allows one to  switch between the dual formulations by imposing appropriate gauge-fixing conditions.

\section{Examples}\label{Sec:4}

In this section, we exemplify the general prescriptions of previous two sections by constructing the dual formulations and ``parent actions" for massive spin one and two, and massless spin two.

\subsection{Massive spin $1$.}
Consider Proca action in $d$-dimensional Minkowski space,
\begin{equation}\label{S-Proca}
\displaystyle \mathcal{S}=\int d^dx\,\mathcal{L}\,, \quad \mathcal{L}=-\,\frac{1}{4}F_{\mu\nu}F^{\mu\nu}+\frac{m^2}{2}A_\mu A^\mu\,, \quad F_{\mu\nu}=\partial_\mu A_\nu-\partial_\nu A_\mu\,,
\end{equation}
cf. (\ref{S}). Here, operator $\hat{M}$ reads
\begin{equation}
\displaystyle \hat{M}_{\mu\nu}=\eta_{\mu\nu}(\square+m^2)-\partial_\mu\partial_\nu\,.
\end{equation}
Corresponding EoMs (\ref{EoMs}) are the Proca equations:
\begin{equation}\label{EoMs-spin-1}
\displaystyle E_\mu\equiv\frac{\delta S}{\delta A^\mu}=(\square+m^2)A_\mu-\partial_\mu\partial_\nu A^\nu=0\,.
\end{equation}
The equations are not involutive as they admit the first order consequence. The involutive closure (cf. (\ref{Inv})) reads
\begin{equation}\label{Inv-spin-1}
\displaystyle E_\mu=0\,, \quad \tau=0\,,
\end{equation}
where $E_\mu$ is defined by (\ref{EoMs-spin-1}), and
\begin{equation}\label{tau-spin-1}
\displaystyle \tau\equiv \partial^\mu E_\mu=m^2\partial_\nu A^\nu\approx 0\,,
\end{equation}
i.e. $\Gamma^{\dagger\mu}=\partial^\mu$, cf. (\ref{diffcons}).
By construction, there exist gauge identities (\ref{GI-closure}),
\begin{equation}
\displaystyle \partial^\mu{}E_\mu-\tau \equiv 0\,.
\end{equation}
The consequence $\tau$ (\ref{tau-spin-1}), being considered on its own, admits gauge symmetry transformation,
\begin{equation}\label{gtr-spin-1}
\delta^{(0)}_\zeta A^\mu=\partial_\nu \zeta^{\mu\nu}\,, \quad \zeta^{\mu\nu}=-\,\zeta^{\nu\mu}\,,
\end{equation}
cf. (\ref{gtr-0}), where
\begin{equation}\label{rho-0-spin-1}
\displaystyle \hat{\rho}^{\mu}{}_{\alpha\beta}=\frac{1}{2}(\delta^\mu_\alpha\partial_\beta-\delta^\mu_\beta\partial_\alpha)\,.
\end{equation}
Gauge transformation (\ref{gtr-spin-1}) is a general solution to the transversality equation (\ref{tau-spin-1}) being the first order consequence of Proca equations.
Transformations (\ref{gtr-spin-1}) are reducible,
\begin{eqnarray}\label{gtr-red-spin-1}
\displaystyle &\delta^{(1)}_{\zeta_1}\zeta^{\mu\nu}=\partial_\lambda\zeta_1^{\mu\nu\lambda}\,, \quad \zeta_1^{\mu\nu\lambda}=\zeta_1^{[\mu\nu\lambda]}\,;\\\label{gtr-red-spin-k}
\displaystyle &\delta^{(k)}_{\zeta_k}\zeta_{k-1}^{\mu\nu\lambda_1\ldots\lambda_{k-1}}=\partial_{\lambda_k}\zeta_k^{\mu\nu\lambda_1\ldots\lambda_k}\,, \quad \zeta_k^{\mu\nu\lambda_1\ldots\lambda_k}=\zeta_k^{[\mu\nu\lambda_1\ldots\lambda_k]}\,, \quad k=2,\ldots,d-2\,,
\end{eqnarray}
cf. (\ref{gtr-1})--(\ref{gtr-k}), where
\begin{eqnarray}\label{rho-1-spin-1}
\displaystyle &\hat{\rho}^{\mu\nu}{}_{\alpha\beta\gamma}=\displaystyle \frac{1}{3}(\delta^\mu_{[\alpha}\delta^\nu_{\beta]}\partial_\gamma+\delta^\mu_{[\beta}\delta^\nu_{\gamma]}\partial_\alpha+\delta^\mu_{[\gamma}\delta^\nu_{\alpha]}\partial_\beta)\,;\\[2mm]
\displaystyle &\hat{\rho}^{\mu\nu\lambda_1\ldots\lambda_{k-1}}{}_{\alpha\beta\gamma_1\ldots\gamma_k}=
\delta^\mu_{[\alpha}\delta^\nu_\beta\delta^{\lambda_1}_{\gamma_1}\ldots\delta^{\lambda_{k-1}}_{\gamma_{k-1}}\partial_{\gamma_k]}\,, \quad k=2,\ldots,d-2\,.
\end{eqnarray}
The sequence of reducible gauge transformations (\ref{gtr-spin-1}), (\ref{gtr-red-spin-1})--(\ref{gtr-red-spin-k}) allows one to compute the DoF number of equation (\ref{tau-spin-1}), and see that it is topological indeed.

According to the general prescription of Section \ref{Sec:2}, to construct the dual formulation we have to add to the involutive closure (\ref{Inv-spin-1}), (\ref{tau-spin-1}) further differential consequences (\ref{TA}) being  variations of the original action (\ref{S-Proca}) with respect to the gauge transformations (\ref{gtr-spin-1}) of topological equation (\ref{tau-spin-1}). These consequences read
\begin{equation}\label{T-spin1}
\displaystyle \mathcal{T}_{\alpha\beta}\equiv \frac{1}{2}(\partial_\alpha E_\beta-\partial_\beta E_\alpha)=\frac{1}{2}(\square+m^2)F_{\alpha\beta}\approx 0\,,
\end{equation}
cf. (\ref{TA}), where $\hat{\rho}^\dagger_{\alpha\beta}{}^\mu$ is a conjugate to $\hat{\rho}^\mu{}_{\alpha\beta}$ (\ref{rho-0-spin-1}).
By construction, there exist gauge identities (\ref{GI-TA}),
\begin{equation}
\displaystyle \mathcal{T}_{\alpha\beta}-\partial_{[\alpha}E_{\beta]}\equiv0\,.
\end{equation}
The theory admit further identities between $\mathcal{T}_{\alpha\beta}$,
\begin{equation}\label{T-spin-1}
\displaystyle -\,\frac{1}{3}(\partial_\gamma\mathcal{T}_{\alpha\beta}+\partial_\alpha\mathcal{T}_{\beta\gamma}+\partial_\beta\mathcal{T}_{\gamma\alpha})\equiv0\,,
\end{equation}
cf. (\ref{TA-1-2}), where $\hat{\rho}^\dagger_{\alpha\beta\gamma}{}^{\mu\nu}$ is a conjugate to $\hat{\rho}{}^{\mu\nu}{}_{\alpha\beta\gamma}$ (\ref{rho-1-spin-1}), as
\begin{equation}
\displaystyle \hat{\rho}^\dagger_{\alpha'\beta'\gamma'}{}^{\alpha\beta}\hat{\rho}^\dagger_{\alpha\beta}{}^\mu=\frac{1}{6}(\delta^\mu_{\alpha'}\partial_{[\gamma'}\partial_{\beta']}+\delta^\mu_{\beta'}\partial_{[\alpha'}\partial_{\gamma']}+
\delta^\mu_{\gamma'}\partial_{[\beta'}\partial_{\alpha']})\equiv0\,.
\end{equation}
These identities admit further reducibility,
\begin{equation}
\displaystyle \hat{\rho}^\dagger_{\mu\nu\lambda\rho}{}^{\alpha\beta\gamma}\hat{\rho}^\dagger{}_{\alpha\beta\gamma}{}^{\alpha'\beta'}\equiv0\,, \quad
\hat{\rho}^\dagger_{\mu\nu\lambda\rho}{}^{\alpha\beta\gamma}=-\,\delta_{[\mu}^\alpha\delta_\nu^\beta\delta_\lambda^\gamma\partial_{\rho]}\,;
\end{equation}
\begin{equation}
\displaystyle \hat{\rho}^\dagger_{\mu\nu\lambda_1\ldots\lambda_k}{}^{\alpha\beta\gamma_1\ldots\gamma_{k-1}}\hat{\rho}^\dagger_{\alpha\beta\gamma_1\ldots\gamma_{k-1}}{}^{\alpha'\beta'\gamma'_1\ldots\gamma'_{k-2}}\equiv0\,, \quad
\hat{\rho}^\dagger_{\mu\nu\lambda_1\ldots\lambda_k}{}^{\alpha\beta\gamma_1\ldots\gamma_{k-1}}=-\,\delta^\alpha_{[\mu}\delta^\beta_\nu\delta^{\gamma_1}_{\lambda_1}\ldots\delta^{\gamma_{k-1}}_{\lambda_{k-1}}\partial_{\gamma_k]}\,,
\end{equation}
where $k=3,\dots,d-2$, cf. (\ref{TA-2})--(\ref{TA-k}).

The involutive closure (\ref{Inv1}) for the Proca model reads
\begin{equation}\label{Inv1-spin-1}
\displaystyle E_\mu=0\,, \quad \tau=0\,, \quad \mathcal{T}_{\alpha\beta}=0\,,
\end{equation}
where $E_\mu$, $\tau$, $\mathcal{T}_{\alpha\beta}$ are defined by (\ref{EoMs-spin-1}), (\ref{tau-spin-1}) and (\ref{T-spin1}), respectively.

Dual equations for (\ref{EoMs-spin-1}), being constructed by the recipe (\ref{dual}), read
\begin{equation}\label{dual-spin-1}
\overline{E}_{\mu}\equiv(\square+m^2)\partial^\nu\omega_{\mu\nu}=0\,,
\quad \omega^{\mu\nu}=-\,\omega^{\nu\mu}\,.
\end{equation}
Gauge transformations for dual equations (\ref{dual-spin-1}) follow from the reducible transformations (\ref{gtr-red-spin-1})--(\ref{gtr-red-spin-k}),
\begin{eqnarray}
\displaystyle &\delta^{(0)}_{\xi_1}\omega^{\mu\nu}=\partial_\lambda\xi_1^{\mu\nu\lambda}\,;\\
\displaystyle &\delta^{(k-1)}_{\xi_k}\xi_{k-1}^{\mu\nu\lambda_1\ldots\lambda_{k-1}}=\partial_{\lambda_k}\xi_k^{\mu\nu\lambda_1\ldots\lambda_{k}}\,,
\end{eqnarray}
where $k=2,\ldots, d-2$, and the gauge parameters are arbitrary totally antisymmetric tensors,
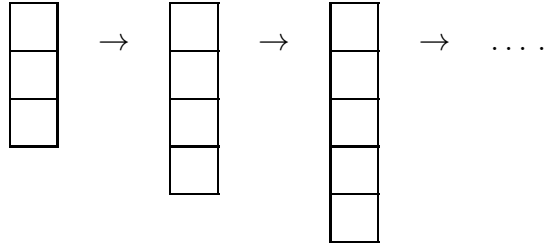
\begin{equation*}
\begin{ytableau}
\phantom{1} \\ \phantom{1} \\ \phantom{1} \
\end{ytableau}
\quad \rightarrow \quad
\begin{ytableau}
\phantom{1} \\ \phantom{1} \\ \phantom{1} \\ \phantom{1} \
\end{ytableau}
\quad \rightarrow \quad
\begin{ytableau}
\phantom{1} \\ \phantom{1} \\ \phantom{1} \\ \phantom{1} \\ \phantom{1} \
\end{ytableau}
\quad \rightarrow \quad \ldots \,.
\end{equation*}
Dual equations (\ref{dual-spin-1}) should  enjoy gauge identities  (\ref{GI-Dual2}) by construction. These identities read
\begin{equation}\label{GI-Dual2-spin1}
\displaystyle -\,\partial^\mu\overline{E}_\mu\equiv0\,.
\end{equation}

Once we have got the involutive closure (\ref{Inv1-spin-1}) (cf. (\ref{Inv1})) for the Proca theory, and dual equations (\ref{dual-spin-1}) for the potentials $\omega^{\mu\nu}$  of the original fields $A^\mu$, we can proceed to constructing the ``parent action" following the general procedure of Section \ref{Sec:3}.
The Stueckelberg action (\ref{SSt}) for massive spin one reads
\begin{equation}\label{SSt-spin-1}
\displaystyle \mathcal{S}[A,\sigma,\omega]=\int d^dx\, \mathfrak{L}\,, \quad \mathfrak{L}=\frac{1}{2}\big(A^\mu-\partial^\mu\sigma+\partial_\lambda\omega^{\mu\lambda}\big)\big[(\square+m^2)(A_\mu+\partial^\nu\omega_{\mu\nu})-\partial_\mu(\partial_\nu A^\nu+m^2\sigma)\big]\,,
\end{equation}
where $\sigma$ and $\omega^{\mu\nu}=-\,\omega^{\nu\mu}$ play the role of Stueckelberg fields. The corresponding EoMs (\ref{EoMStphi})--(\ref{EoMStomega}) take the form
\begin{eqnarray}
\displaystyle \frac{\delta \mathcal{S}}{\delta A^\mu}\equiv(\square+m^2)(A_\mu+\partial^\nu\omega_{\mu\nu})-\partial_\mu(\partial_\nu A^\nu+m^2\sigma)=0\,;\\
\displaystyle \frac{\delta \mathcal{S}}{\delta \sigma}\equiv m^2(\partial_\mu A^\mu-\square\sigma)=0\,;\\
\displaystyle \frac{\delta \mathcal{S}}{\delta \omega^{\alpha\beta}}\equiv\frac{1}{2}(\square+m^2)\big[\partial_\alpha(A_\beta+\partial^\gamma\omega_{\beta\gamma})-\partial_\beta(A_\alpha+\partial^\gamma\omega_{\alpha\gamma})\big]=0\,.
\end{eqnarray}

The Stueckelberg action (\ref{SSt-spin-1}) is invariant under gauge transformations (cf. (\ref{StGS}), (\ref{StGS-reduce-1})--(\ref{StGS-reduce-k}))
\begin{equation}
\displaystyle \delta^{(0)}A^\mu=\partial^\mu\mathcal{E}-\partial_\nu\mathcal{E}^{\mu\nu}\,, \quad \delta^{(0)}\sigma=\mathcal{E}\,, \quad \delta^{(0)}\omega^{\mu\nu}=\mathcal{E}^{\mu\nu}-\partial_\lambda\xi^{\mu\nu\lambda}\,;
\end{equation}
\begin{equation}
\displaystyle \delta^{(1)}\mathcal{E}^{\mu\nu}=\partial_\lambda\mathcal{E}^{\mu\nu\lambda}\,, \quad \delta^{(1)}\mathcal{E}=0\,, \quad \delta^{(1)}\xi^{\mu\nu\lambda}=\mathcal{E}^{\mu\nu\lambda}-\partial_\rho\xi^{\mu\nu\lambda\rho}\,;
\end{equation}
\begin{equation}
\displaystyle \delta^{(k)}\mathcal{E}^{\mu\nu\lambda_1\ldots\lambda_{k-1}}=\partial_{\lambda_k}\mathcal{E}^{\mu\nu\lambda_1\ldots\lambda_k}\,, \quad
\delta^{(k)}\xi^{\mu\nu\lambda_1\ldots\lambda_k}=\mathcal{E}^{\mu\nu\lambda_1\ldots\lambda_k}-\partial_{\lambda_{k+1}}\xi^{\mu\nu\lambda_1\ldots\lambda_{k+1}}\,,
\end{equation}
where $k=2,\ldots,d-3$;
\begin{equation}
\displaystyle \delta^{(d-2)}\mathcal{E}^{\mu\nu\lambda_1\ldots\lambda_{d-3}}=\partial_{\lambda_{d-2}}\mathcal{E}^{\mu\nu\lambda_1\ldots\lambda_{d-2}}\,, \quad \delta^{(d-2)}\xi^{\mu\nu\lambda_1\ldots\lambda_{d-2}}=\mathcal{E}^{\mu\nu\lambda_1\ldots\lambda_{d-2}}\,.
\end{equation}

Gauge-fixing condition
\begin{equation}
\displaystyle \sigma=0\,, \quad \omega^{\mu\nu}=0\,,
\end{equation}
reproduce original theory (\ref{EoMs-spin-1}), while another admissible gauge
\begin{equation}
\displaystyle A^\mu=0\,, \quad \sigma=0\,, \quad -\frac{1}{3}(\partial^\lambda\omega^{\mu\nu}+\partial^\mu\omega^{\nu\lambda}+\partial^\nu\omega^{\lambda\mu})=0\,,
\end{equation}
i.e. $\hat{\rho}^\dagger{}^{\mu\nu\lambda}{}_{\alpha\beta}=-\,\delta^{[\mu}_\alpha\delta^\nu_\beta\partial^{\lambda]}$, cf. (\ref{gauge2}), leads to dual formulation (\ref{dual-spin-1}).

The dualisation of massive spin one constructed by the proposed general procedure has a simple outcome: the divergence of the antisymmetric tensor $\omega^{\nu\mu}$, being the ``potential" for massive vector (cf. (\ref{gtr-spin-1})), has to obey Klein-Gordon equation (\ref{dual-spin-1}).
Unlike the Proca equations, the dual ones are homogeneous and involutive --- they are all of the third order, and they do not admit any lower order consequence. Hence, to control the consistency of interactions, one has to care only about consistent deformation of gauge algebra. The dual equations (\ref{dual-spin-1}) are not Lagrangian as such though they can be cast into Lagrangian framework by constructing the parent action (\ref{SSt-spin-1}). This action involves both massive vector $A^\mu$, and its ``potential" $\omega^{\mu\nu}$. Lagrangian equations for the parent action are also involutive, unlike Proca ones. From the perspective of the parent action, one can switch between these two dual theories by imposing different gauge-fixing conditions.

\subsection{Massive spin $2$.}
Consider a theory of massive spin two represented by symmetric second rank tensor $h^{\mu\nu}$ in $d=4$ Minkowski space,
\begin{equation}\label{S-spin-2m}
\displaystyle S=\int d^4x\,\mathcal{L}\,, \quad \mathcal{L}=\frac{1}{2}\big(h\square h-2h^{\mu\nu}\partial_\mu\partial_\nu h-h^{\mu\nu}\square h_{\mu\nu}+2h^{\mu\nu}\partial_\nu\partial^\lambda h_{\mu\lambda}\big)-\frac{m^2}{2}\big(h^{\mu\nu} h_{\mu\nu}-h^2\big)\,,
\end{equation}
where $ h\equiv\eta_{\mu\nu}h^{\mu\nu}$. The operator $\hat{M}_{\mu\nu\lambda\rho}$ (cf. (\ref{S})) reads
\begin{equation}
\begin{array}{c}
\displaystyle \hat{M}_{\mu\nu\lambda\rho}=\big(\eta_{\mu\nu}\eta_{\lambda\rho}-\frac{1}{2}(\eta_{\mu\lambda}\eta_{\nu\rho}+\eta_{\nu\lambda}\eta_{\mu\rho})\big)(\square+m^2)
-\eta_{\lambda\rho}\partial_\mu\partial_\nu
-\eta_{\mu\nu}\partial_\lambda\partial_\rho\\[2mm]
\displaystyle +\,\frac{1}{2}\big(\eta_{\mu\lambda}\partial_\nu\partial_\rho+\eta_{\nu\lambda}\partial_\mu\partial_\rho+\eta_{\mu\rho}\partial_\nu\partial_\lambda+\eta_{\nu\rho}\partial_\mu\partial_\lambda\big)\,.
 \end{array}
\end{equation}
Corresponding Lagrangian equations (\ref{EoMs})
\begin{equation}\label{EoMs-spin-2m}
\displaystyle E_{\mu\nu}\equiv\frac{\delta S}{\delta h^{\mu\nu}}=(\square+m^2)(\eta_{\mu\nu}h-h_{\mu\nu})-\partial_\mu\partial_\nu h-\eta_{\mu\nu}\partial_\lambda\partial_\rho h^{\lambda\rho}+\partial_\mu\partial^\lambda h_{\nu\lambda}+\partial_\nu\partial^\lambda h_{\mu\lambda}=0
\end{equation}
are of second order. These equations admit zero and first order consequences:
\begin{eqnarray}\label{tau-spin-2m}
&\displaystyle \tau\equiv\frac{1}{d-1}\eta^{\mu\nu}E_{\mu\nu}-\frac{d-2}{d-1}m^{-2}\partial^{\mu}\partial^{\nu}E_{\mu\nu}=m^2h\approx0\,;\\[2mm]
\label{taualpha-spin-2m}
&\displaystyle \widetilde{\tau}_\alpha\equiv-\,\frac{d-2}{d}m^{-2}\partial_\alpha\partial^\mu\partial^\nu E_{\mu\nu}-\partial^\nu E_{\mu\nu}+\frac{1}{d}\eta^{\mu\nu}\partial_\alpha E_{\mu\nu}=m^2 \partial^\beta \widetilde{h}_{\alpha\beta}\approx 0\,,
\end{eqnarray}
where $\widetilde{h}{}^{\alpha\beta}$ is the traceless part of ${h}{}^{\alpha\beta}$, $\displaystyle \widetilde{h}{}^{\alpha\beta}=h^{\alpha\beta}-\frac{1}{d}\eta^{\alpha\beta}h$,  $\,\,\eta_{\eta\beta}\widetilde{h}{}^{\alpha\beta}\equiv 0$.
Comparing to the spin one case, besides the transversality condition (\ref{taualpha-spin-2m}), we have algebraic equation (\ref{tau-spin-2m}).

For the massive spin two, the generators of the lower order consequences (\ref{tau-spin-2m}), (\ref{taualpha-spin-2m}) read
\begin{eqnarray}
&\displaystyle \hat{\Gamma}^\dagger{}^{\mu\nu}=\frac{1}{d-1}\big(\eta^{\mu\nu}-(d-2)m^{-2}\partial^{\mu}\partial^{\nu}\big)\,;
\\[2mm]
&\displaystyle \hat{\widetilde{\Gamma}}{}^\dagger_\alpha{}^{\mu\nu}=-\,\frac{d-2}{d}m^{-2}\partial_\alpha\partial^\mu\partial^\nu-\frac{1}{2}\big(\delta^\mu_\alpha\partial^\nu+\delta^\nu_\alpha\partial^\mu-\frac{2}{d}\eta^{\mu\nu}\partial_\alpha\big)\,,
\end{eqnarray}
cf. (\ref{diffcons}). Involutively closed system (\ref{Inv}) for massive spin two field takes the form
\begin{equation}\label{Inv-spin-2m}
\displaystyle E_{\mu\nu}=0\,, \quad \tau=0\,, \quad \widetilde{\tau}_\alpha=0\,,
\end{equation}
where $E_{\mu\nu}$, $\tau$ and $\widetilde{\tau}_{\alpha}$ are defined by (\ref{EoMs-spin-2m}), (\ref{tau-spin-2m}) and (\ref{taualpha-spin-2m}), respectively. By construction, there exist gauge identities (\ref{GI-closure}),
\begin{eqnarray}
&\displaystyle \frac{1}{d-1}\eta^{\mu\nu}E_{\mu\nu}-\frac{d-2}{d-1}m^{-2}\partial^{\mu}\partial^{\nu}E_{\mu\nu}-\tau \equiv 0\,;\\
&\displaystyle -\,\frac{d-2}{d}m^{-2}
\partial_\alpha\partial^\mu\partial^\nu E_{\mu\nu}-\partial^\nu E_{\mu\nu}+\frac{1}{d}\eta^{\mu\nu}\partial_\alpha E_{\mu\nu}-\widetilde{\tau}_\alpha\equiv0\,.
\end{eqnarray}
The consequences $\tau$,  $\widetilde{\tau}_\alpha$, being considered on their own, admit gauge symmetry transformation,
\begin{equation}\label{gtr-spin-2m}
\delta^{(0)}_\zeta h^{\mu\nu}=\partial_\lambda\partial_\rho\widetilde{\zeta}{}^{\mu\nu\lambda\rho}\,,
\end{equation}
where $\widetilde{\zeta}{}^{\mu\nu\lambda\rho}$ is a traceless tensor with window symmetry type, described by the Young diagram
\begin{equation}
\displaystyle
\begin{ytableau}
\phantom{1} & \phantom{1} \\ \phantom{1} & \phantom{1}\
\end{ytableau}\,:
\quad \widetilde{\zeta}{}^{(\mu\nu)\lambda\rho}=\widetilde{\zeta}{}^{\mu\nu\lambda\rho}\,,
\quad \widetilde{\zeta}{}^{\mu\nu(\lambda\rho)}=\widetilde{\zeta}{}^{\mu\nu\lambda\rho}\,, \quad \widetilde{\zeta}{}^{(\mu\nu\lambda)\rho}=0\,, \quad
\eta_{\mu\nu}\widetilde{\zeta}{}^{\mu\nu\lambda\rho}=0\,,
\end{equation}
cf. (\ref{gtr-0}). Here,
\begin{equation}\label{rho-0-spin-2m}
\begin{array}{c}
\displaystyle \hat{\rho}{}^{\mu\nu}{}_{\alpha\beta\gamma\delta}=
\frac{1}{3}\Big(2(\delta^\mu_{(\alpha}\delta^\nu_{\beta)}-\frac{1}{d-1}\eta^{\mu\nu}\eta_{\alpha\beta})\partial_\gamma\partial_\delta-\frac{1}{2}(\delta^\mu_{(\beta}\delta^\nu_{\gamma)}-\frac{1}{d-1}\eta^{\mu\nu}\eta_{\beta\gamma})\partial_\alpha\partial_\delta
\\[2mm]
\displaystyle -\,\frac{1}{2}(\delta^\mu_{(\beta}\delta^\nu_{\delta)}-\frac{1}{d-1}
\eta^{\mu\nu}\eta_{\beta\delta})\partial_\alpha\partial_\gamma-\frac{1}{2}(\delta^\mu_{(\alpha}\delta^\nu_{\gamma)}-\frac{1}{d-1}\eta^{\mu\nu}\eta_{\beta\gamma})\partial_\beta\partial_\delta
\\[2mm]
\displaystyle -\,\frac{1}{2}(\delta^\mu_{(\alpha}\delta^\nu_{\delta)}-\frac{1}{d-1}\eta^{\mu\nu}\eta_{\alpha\delta})\partial_\beta\partial_\gamma+\frac{1}{d-1}
\eta_{\alpha\beta}\delta^{(\mu}_\gamma\partial^{\nu)}\partial_\delta+\frac{1}{d-1}
\eta_{\alpha\beta}\delta^{(\mu}_\delta\partial^{\nu)}\partial_\gamma
\\[2mm]
\displaystyle -\,\frac{1}{2}\frac{1}{d-1}\eta_{\beta\gamma}\delta^{(\mu}_\alpha\partial^{\nu)}\partial_\delta-\frac{1}{2}\frac{1}{d-1}\eta_{\beta\delta}\delta^{(\mu}_\alpha\partial^{\nu)}\partial_\gamma-\frac{1}{2}\frac{1}{d-1}
\eta_{\alpha\gamma}\delta^{(\mu}_\beta\partial^{\nu)}\partial_\delta-\frac{1}{2}
\frac{1}{d-1}\eta_{\alpha\delta}\delta^{(\mu}_\beta\partial^{\nu)}\partial_\gamma
\Big)\,.
\end{array}
\end{equation}
Gauge transformation (\ref{gtr-spin-2m}) is a general solution to the transversality equation for symmetric traceless tensor\footnote{See in \cite{Francia:2013sca}.} (\ref{taualpha-spin-2m}), being a first order consequence of EoMs (\ref{EoMs-spin-2m}). These transformations are reducible,
\begin{equation}\label{gtr-red-1-spin-2m}
\begin{array}{c}
\displaystyle \delta^{(1)}_{\zeta_1}\widetilde{\zeta}{}^{\mu\nu\lambda\rho}=\partial^\lambda\partial_\sigma\widetilde{\zeta}{}_1^{\mu\nu\rho\sigma}+\partial^\rho\partial_\sigma\widetilde{\zeta}{}_1^{\mu\nu\lambda\sigma}-\partial^\mu\partial_\sigma\widetilde{\zeta}{}_1^{\nu\lambda\rho\sigma}-\partial^\mu\partial_\sigma\widetilde{\zeta}{}_1^{\nu\rho\lambda\sigma}-\partial^\nu\partial_\sigma\widetilde{\zeta}{}_1^{\mu\lambda\rho\sigma}-\partial^\nu\partial_\sigma\widetilde{\zeta}{}_1^{\mu\rho\lambda\sigma}\,,
\end{array}
\end{equation}
where $\widetilde{\zeta}{}_1^{\mu\nu\lambda\rho}$ is a traceless tensor with the hook symmetry,
\begin{equation}
\displaystyle
\begin{ytableau}
\phantom{1} & \phantom{1} \\ \phantom{1} \\ \phantom{1} \
\end{ytableau}\,:
\quad \widetilde{\zeta}{}_1^{(\mu\nu)\lambda\rho}=\widetilde{\zeta}{}_1^{\mu\nu\lambda\rho}\,,
\quad \widetilde{\zeta}{}_1^{\mu\nu(\lambda\rho)}=0\,,
\quad \widetilde{\zeta}{}_1^{(\mu\nu\lambda)\rho}=0\,, \quad
\eta_{\mu\nu}\widetilde{\zeta}{}_1^{\mu\nu\lambda\rho}=0\,;
\end{equation}
\begin{equation}\label{gtr-red-2-spin-2m}
\begin{array}{c}
\displaystyle \delta^{(2)}_{\zeta_2}\widetilde{\zeta}{}_1^{\mu\nu\lambda\rho}=\partial^\mu\zeta{}_2^{\nu\lambda\rho}+\partial^\nu\zeta{}_2^{\mu\lambda\rho}-\frac{1}{d-2}\big(2\eta^{\mu\nu}\partial_\sigma\zeta{}_2^{\sigma\lambda\rho}
+\eta^{\mu\lambda}\partial_\sigma\zeta{}_2^{\nu\sigma\rho}\\[2mm]
\displaystyle +\,\eta^{\nu\lambda}\partial_\sigma\zeta{}_2^{\mu\sigma\rho}+\eta^{\mu\rho}\partial_\sigma\zeta{}_2^{\nu\lambda\sigma}+\eta^{\nu\rho}\partial_\sigma\zeta{}_2^{\mu\lambda\sigma}\big)\,,
\end{array}
\end{equation}
where $\zeta{}_2^{\mu\nu\lambda}$ is a totally antisymmetric tensor,
\begin{equation}
\displaystyle
\begin{ytableau}
\phantom{1}  \\ \phantom{1} \\ \phantom{1} \
\end{ytableau}\,:
\quad \zeta{}_2^{[\mu\nu\lambda]}=\zeta{}_2^{\mu\nu\lambda}\,,
\end{equation}
cf. (\ref{gtr-1})--(\ref{gtr-k}). Here,
\begin{equation}\label{rho-1-spin-2m}
\begin{array}{c}
\displaystyle \hat{\rho}^{\mu\nu\lambda\rho}{}_{\alpha\beta\gamma\delta}
=\frac{1}{3}\Big(2\delta^\mu_{(\alpha}\delta^\nu_{\beta)}(\delta^\lambda_{[\gamma}\partial_{\delta]}\partial^\rho+\delta^\rho_{[\gamma}\partial_{\delta]}\partial^\lambda)-(\delta^\nu_{(\alpha}\delta^\lambda_{\beta)}\partial^\rho+\delta^\nu_{(\alpha}\delta^\rho_{\beta)}\partial^\lambda)\delta^\mu_{[\gamma}\partial_{\delta]}\\[2mm]
\displaystyle -\,(\delta^\mu_{(\alpha}\delta^\lambda_{\beta)}\partial^\rho+\delta^\mu_{(\alpha}\delta^\rho_{\beta)}\partial^\lambda)\delta^\nu_{[\gamma}\partial_{\delta]}-\frac{1}{d-1}\eta^{\mu\nu}\big(2\eta_{\alpha\beta}(\delta^\lambda_{[\gamma}\partial_{\delta]}\partial^\rho+\delta^\rho_{[\gamma}\partial_{\delta]}\partial^\rho)-\eta_{\beta[\gamma}\partial_{\delta]}(\delta^\lambda_\alpha\partial^\rho+\delta^\rho_\alpha\partial^\lambda)
\\[2mm]
\displaystyle -\,\eta_{\alpha[\gamma}\partial_{\delta]}(\delta^\lambda_\beta\partial^\rho+\delta^\rho_\beta\partial^\lambda)\big)+\frac{1}{2}\frac{1}{d-1}(2\eta_{\alpha\beta}\delta^\nu_{[\gamma}-\delta^\nu_\alpha\eta_{\beta[\gamma}-\delta^\nu_\beta\eta_{\alpha[\gamma})\partial_{\delta]}(\eta^{\lambda\mu}\partial^\rho+\eta^{\rho\mu}\partial^\lambda)
\\[2mm]
\displaystyle +\,\frac{1}{2}\frac{1}{d-1}(2\eta_{\alpha\beta}\delta^\mu_{[\gamma}-\delta^\mu_\alpha\eta_{\beta[\gamma}-\delta^\mu_\beta\eta_{\alpha[\gamma})\partial_{\delta]}(\eta^{\nu\lambda}\partial^\rho+\eta^{\nu\rho}\partial^\lambda)+\big[2\delta^\rho_{(\alpha}\delta^\lambda_{\beta)}\delta^\nu_{[\gamma}-\delta^\nu_{(\alpha}\delta^\rho_{\beta)}\delta^\lambda_{[\gamma}
\\[2mm]
\displaystyle -\,\delta^\nu_{(\alpha}\delta^\lambda_{\beta)}
\delta^\rho_{[\gamma}+\frac{1}{2}\frac{1}{d-1}\big(
\eta^{\nu\rho}(2\eta_{\alpha\beta}\delta^\lambda_{[\gamma}-\delta^\lambda_\alpha\eta_{\beta[\gamma}-\delta^\lambda_\beta\eta_{\alpha[\gamma})
+\eta^{\lambda\nu}(2\eta_{\alpha\beta}\delta^\rho_{[\gamma}-\delta^\rho_\alpha\eta_{\beta[\gamma}-\delta^\rho_\beta\eta_{\alpha[\gamma})
\\[2mm]
\displaystyle -\,2\eta^{\rho\lambda}(2\eta_{\alpha\beta}\delta^\nu_{[\gamma}-\delta^\nu_\alpha\eta_{\beta[\gamma}-\delta^\nu_\beta\eta_{\alpha[\gamma})\big)\big]\partial_{\delta]}\partial^\mu+\big[2\delta^\rho_{(\alpha}\delta^\lambda_{\beta)}\delta^\mu_{[\gamma}-\delta^\mu_{(\alpha}\delta^\rho_{\beta)}\delta^\lambda_{[\gamma}-\delta^\mu_{(\alpha}\delta^\lambda_{\beta)}\delta^\rho_{[\gamma}
\\[2mm]
\displaystyle +\,\frac{1}{2}\frac{1}{d-1}\big(
\eta^{\mu\rho}(2\eta_{\alpha\beta}\delta^\lambda_{[\gamma}-\delta^\lambda_\alpha\eta_{\beta[\gamma}-\delta^\lambda_\beta\eta_{\alpha[\gamma})+\eta^{\lambda\mu}(2\eta_{\alpha\beta}\delta^\rho_{[\gamma}-\delta^\rho_\alpha\eta_{\beta[\gamma}-\delta^\rho_\beta\eta_{\alpha[\gamma})
\\[2mm]
\displaystyle -\,2\eta^{\rho\lambda}(2\eta_{\alpha\beta}\delta^\mu_{[\gamma}-\delta^\mu_\alpha\eta_{\beta[\gamma}-\delta^\mu_\beta\eta_{\alpha[\gamma})\big)\big]\partial_{\delta]}\partial^\nu\Big)\,;
\end{array}
\end{equation}
\begin{equation}\label{rho-2-spin-2m}
\begin{array}{c}
\displaystyle \hat{\rho}^{\mu\nu\lambda\rho}{}_{\alpha\beta\gamma}
=\frac{1}{3}\Big(
\big(\delta^\nu_\alpha\delta^\lambda_{[\beta}\delta^\rho_{\gamma]}+
\delta^\nu_\beta\delta^\lambda_{[\gamma}\delta^\rho_{\alpha]}+
\delta^\nu_\gamma\delta^\lambda_{[\alpha}\delta^\rho_{\beta]}\big)\partial^\mu+\big(\delta^\mu_\alpha\delta^\lambda_{[\beta}\delta^\rho_{\gamma]}+
\delta^\mu_\beta\delta^\lambda_{[\gamma}\delta^\rho_{\alpha]}+
\delta^\mu_\gamma\delta^\lambda_{[\alpha}\delta^\rho_{\beta]}\big)\partial^\nu
\\[2mm]
\displaystyle -\,\frac{1}{d-2}\big[2\eta^{\mu\nu}\big(
\delta^\lambda_{[\beta}\delta^\rho_{\gamma]}\partial_\alpha+
\delta^\lambda_{[\gamma}\delta^\rho_{\alpha]}\partial_\beta+
\delta^\lambda_{[\alpha}\delta^\rho_{\beta]}\partial_\gamma\big)
-\eta^{\mu\lambda}\big(\delta^\nu_{[\beta}\delta^\rho_{\gamma]}\partial_\alpha+
\delta^\nu_{[\gamma}\delta^\rho_{\alpha]}\partial_\beta+
\delta^\nu_{[\alpha}\delta^\rho_{\beta]}\partial_\gamma\big)
\\[2mm]
\displaystyle -\,\eta^{\nu\lambda}\big(\delta^\mu_{[\beta}
\delta^\rho_{\gamma]}\partial_\alpha+\delta^\mu_{[\gamma}\delta^\rho_{\alpha]}\partial_\beta+\delta^\mu_{[\alpha}\delta^\rho_{\beta]}\partial_\gamma\big)
+\eta^{\mu\rho}\big(\delta^\nu_{[\beta}\delta^\lambda_{\gamma]}\partial_\alpha+
\delta^\nu_{[\gamma}\delta^\lambda_{\alpha]}\partial_\beta+\delta^\nu_{[\alpha}\delta^\lambda_{\beta]}\partial_\gamma\big)
\\[2mm]
\displaystyle +\,\eta^{\nu\rho}\big(\delta^\mu_{[\beta}\delta^\lambda_{\gamma]}\partial_\alpha+\delta^\mu_{[\gamma}\delta^\lambda_{\alpha]}\partial_\beta
+\delta^\mu_{[\alpha}\delta^\lambda_{\beta]}\partial_\gamma\big)\big]\Big)\,.
\end{array}
\end{equation}
As in the case of massive spin one, the theory described by equations (\ref{tau-spin-2m})--(\ref{taualpha-spin-2m}) with reducible gauge symmetry transformations (\ref{gtr-spin-2m}), (\ref{gtr-red-1-spin-2m}) and (\ref{gtr-red-2-spin-2m}) is topological.

According to the general procedure, we need to add to the involutive closure (\ref{Inv-spin-2m}) the higher order differential consequences, being variations of the original action (\ref{S-spin-2m}) with respect to gauge transformations (\ref{gtr-spin-2m}). These consequences read
\begin{equation}\label{TA-spin-2m}
\begin{array}{c}
\displaystyle \widetilde{\mathcal{T}}_{\alpha\beta\gamma\delta}\equiv\frac{1}{3}(\square+m^2)\Big(2\partial_\gamma\partial_\delta(h_{\alpha\beta}-\frac{1}{d-1}\eta_{\alpha\beta}h)-\frac{1}{2}\partial_\alpha\partial_\delta
(h_{\beta\gamma}-\frac{1}{d-1}\eta_{\beta\gamma}h)\\[2mm]
\displaystyle -\,\frac{1}{2}\partial_\alpha\partial_\gamma(h_{\beta\delta}-
\frac{1}{d-1}\eta_{\beta\delta}h)-\frac{1}{2}\partial_\beta\partial_\delta
(h_{\alpha\gamma}-\frac{1}{d-1}\eta_{\beta\gamma})h
\\[2mm]
\displaystyle -\,\frac{1}{2}\partial_\beta\partial_\gamma(h_{\alpha\delta}
-\frac{1}{d-1}\eta_{\alpha\delta}h)+\frac{1}{2}\frac{1}{d-1}\big(2\eta_{\alpha\beta}(\partial_\delta\partial^\lambda h_{\gamma\lambda}+\partial_\gamma\partial^\lambda h_{\delta\lambda})
\\[2mm]
\displaystyle -\,\eta_{\beta\gamma}\partial_\delta\partial^\lambda h_{\alpha\lambda}-\eta_{\beta\delta}\partial_\gamma\partial^\lambda h_{\alpha\lambda}-\eta_{\alpha\gamma}\partial_\delta\partial^\lambda h_{\beta\lambda}-\eta_{\alpha\delta}\partial_\gamma\partial^\lambda h_{\beta\lambda}\big)\Big)\approx 0\,,
\end{array}
\end{equation}
cf. (\ref{TA}), where $\hat{\rho}^\dagger_{\alpha\beta\gamma\delta}{}^{\mu\nu}$ is a conjugate to $\hat{\rho}^{\mu\nu}{}_{\alpha\beta\gamma\delta}$ (\ref{rho-0-spin-2m}). By construction, $\widetilde{\mathcal{T}}_{\alpha\beta\gamma\delta}$ enjoy gauge identities
\begin{equation}
\begin{array}{c}
\displaystyle \widetilde{\mathcal{T}}_{\alpha\beta\gamma\delta}-\frac{1}{3}\Big(2\partial_\gamma\partial_\delta
(E_{\alpha\beta}-\frac{1}{d-1}\eta_{\alpha\beta}\eta^{\mu\nu}E_{\mu\nu})
-\frac{1}{2}\partial_\alpha\partial_\delta(E_{\beta\gamma}-\frac{1}{d-1}
\eta_{\beta\gamma}\eta^{\mu\nu}E_{\mu\nu})
\\[2mm]
\displaystyle -\,\frac{1}{2}\partial_\alpha\partial_\gamma(E_{\beta\delta}-
\frac{1}{d-1}\eta_{\beta\delta}\eta^{\mu\nu}E_{\mu\nu})-\frac{1}{2}\partial_\beta\partial_\delta(E_{\alpha\gamma}-\frac{1}{d-1}\eta_{\beta\gamma})\eta^{\mu\nu}E_{\mu\nu}
\\[2mm]
\displaystyle -\,\frac{1}{2}\partial_\beta\partial_\gamma(E_{\alpha\delta}
-\frac{1}{d-1}\eta_{\alpha\delta}\eta^{\mu\nu}E_{\mu\nu})+\frac{1}{2}\frac{1}{d-1}\big(2\eta_{\alpha\beta}(\partial_\delta\partial^\lambda E_{\gamma\lambda}+\partial_\gamma\partial^\lambda E_{\delta\lambda})
\\[2mm]
\displaystyle -\,\eta_{\beta\gamma}\partial_\delta\partial^\lambda E_{\alpha\lambda}-\eta_{\beta\delta}\partial_\gamma\partial^\lambda E_{\alpha\lambda}-\eta_{\alpha\gamma}\partial_\delta\partial^\lambda E_{\beta\lambda}-\eta_{\alpha\delta}\partial_\gamma\partial^\lambda E_{\beta\lambda}\big)\Big)\equiv0\,.
\end{array}
\end{equation}
There exist identities between (\ref{TA-spin-2m}):
\begin{equation}\label{T-spin-2m}
\begin{array}{c}
\displaystyle \displaystyle \frac{2}{3}
\Big(\partial^\lambda\big(2(\partial_{[\delta}\widetilde{\mathcal{T}}_{\gamma]\lambda\alpha\beta}-\partial_{[\delta}\widetilde{\mathcal{T}}_{\gamma](\alpha\beta)\lambda})+\partial_\delta(\widetilde{\mathcal{T}}_{\alpha\beta\gamma\lambda}-\widetilde{\mathcal{T}}_{\lambda(\alpha\beta)\gamma})-\partial_\gamma(\widetilde{\mathcal{T}}_{\alpha\beta\delta\lambda}-\widetilde{\mathcal{T}}_{\lambda(\alpha\beta)\delta})\big)
\\[2mm]
\displaystyle +\,\frac{1}{d-1}\eta^{\lambda\rho}\partial^\sigma\big[\eta_{\alpha\beta}\big(2(\partial_{[\delta}\widetilde{\mathcal{T}}_{\gamma]\lambda\rho\sigma}-\partial_{[\delta}\widetilde{\mathcal{T}}_{\gamma]\sigma\lambda\rho})+\partial_\delta\widetilde{\mathcal{T}}_{\sigma\lambda\rho\gamma}-\partial_\gamma\widetilde{\mathcal{T}}_{\sigma\lambda\rho\delta}\big)
\\[2mm]
\displaystyle
-\eta_{\beta[\gamma}\partial_{\delta]}(\widetilde{\mathcal{T}}_{\alpha\lambda\rho\sigma}+\widetilde{\mathcal{T}}_{\sigma\lambda\rho\alpha}-\widetilde{\mathcal{T}}_{\alpha\sigma\lambda\rho})-\eta_{\alpha[\gamma}\partial_{\delta]}(\widetilde{\mathcal{T}}_{\beta\lambda\rho\sigma}+\widetilde{\mathcal{T}}_{\sigma\lambda\rho\beta}-\widetilde{\mathcal{T}}_{\beta\sigma\lambda\rho})\big]\Big)\equiv0\,,
\end{array}
\end{equation}
as
\begin{equation}
\displaystyle \hat{\rho}^\dagger_{\alpha\beta\gamma\delta}{}^{\mu\nu\lambda\rho}\hat{\rho}^\dagger_{\mu\nu\lambda\rho}{}^{\alpha'\beta'}\equiv0\,,
\end{equation}
cf. (\ref{TA-1-1})--(\ref{TA-1-2}), where $\hat{\rho}^\dagger_{\alpha\beta\gamma\delta}{}^{\mu\nu\lambda\rho}$ is a conjugate to $\hat{\rho}{}^{\mu\nu\lambda\rho}{}_{\alpha\beta\gamma\delta}$ (\ref{rho-1-spin-2m}).
The identities (\ref{T-spin-2m}) are further reducible,
\begin{equation}
\displaystyle \hat{\rho}^\dagger_{\alpha\beta\gamma}{}^{\mu\nu\lambda\rho}\hat{\rho}^\dagger{}_{\mu\nu\lambda\rho}{}^{\alpha'\beta'\gamma'\delta'}\equiv0\,,
\end{equation}
cf. (\ref{TA-2}), where $\hat{\rho}^\dagger_{\alpha\beta\gamma}{}^{\mu\nu\lambda\rho}$ is a conjugate to $\hat{\rho}{}^{\mu\nu\lambda\rho}{}_{\alpha\beta\gamma}$ (\ref{rho-2-spin-2m}).

The involutive closure (\ref{Inv1}) for massive spin two reads
\begin{equation}\label{Inv1-spin-2m}
\displaystyle E_{\mu\nu}=0\,, \quad \tau=0\,, \quad \widetilde{\tau}_\alpha=0\,, \quad \widetilde{\mathcal{T}}_{\alpha\beta\gamma\delta}=0\,,
\end{equation}
where $E_{\mu\nu}$, $\tau$, $\widetilde{\tau}_\alpha$ and $\widetilde{\mathcal{T}}_{\alpha\beta\gamma\delta}$ are defined by (\ref{EoMs-spin-2m}), (\ref{tau-spin-2m}), (\ref{taualpha-spin-2m}) and (\ref{TA-spin-2m}), respectively.

Dual equations for (\ref{EoMs-spin-2m}), being constructed by the recipe (\ref{dual}), read
\begin{equation}\label{dual-spin-2m}
\overline{E}_{\mu\nu}\equiv(\square+m^2)\partial_\lambda\partial_\rho\widetilde{\omega}^{\mu\nu\lambda\rho}=0\,,
\end{equation}
where $\widetilde{\omega}{}^{\mu\nu\lambda\rho}$ is a traceless tensor with window symmetry type ,
\begin{equation}\label{omega-spin-2m}
\displaystyle
\begin{ytableau}
\phantom{1} & \phantom{1} \\ \phantom{1}  & \phantom{1}\
\end{ytableau}\,:
\quad \widetilde{\omega}{}^{(\mu\nu)\lambda\rho}=\widetilde{\omega}{}^{\mu\nu\lambda\rho}\,, \quad
\widetilde{\omega}{}^{\mu\nu(\lambda\rho)}=\widetilde{\omega}{}^{\mu\nu\lambda\rho}\,,
\quad \widetilde{\omega}{}^{(\mu\nu\lambda)\rho}=0\,, \quad
\eta_{\mu\nu}\widetilde{\omega}{}^{\mu\nu\lambda\rho}=0\,.
\end{equation}
Gauge symmetry transformations for (\ref{dual-spin-2m}), following from (\ref{gtr-red-1-spin-2m}) and (\ref{gtr-red-2-spin-2m}), read
\begin{equation}
\displaystyle \delta^{(0)}_{\xi_1}\widetilde{\omega}{}^{\mu\nu\lambda\rho}=\partial^\lambda\partial_\sigma\widetilde{\xi}{}_1^{\mu\nu\rho\sigma}+\partial^\rho\partial_\sigma\widetilde{\xi}{}_1^{\mu\nu\lambda\sigma}-\partial^\mu\partial_\sigma\widetilde{\xi}{}_1^{\nu\lambda\rho\sigma}-\partial^\mu\partial_\sigma\widetilde{\xi}{}_1^{\nu\rho\lambda\sigma}-\partial^\nu\partial_\sigma\widetilde{\xi}{}_1^{\mu\lambda\rho\sigma}-\partial^\nu\partial_\sigma\widetilde{\xi}{}_1^{\mu\rho\lambda\sigma}\,;
\end{equation}
\begin{equation}
\begin{array}{c}
\displaystyle \delta^{(1)}_{\xi_2}\widetilde{\xi}{}_1^{\mu\nu\lambda\rho}=\partial^\mu\xi{}_2^{\nu\lambda\rho}+\partial^\nu\xi{}_2^{\mu\lambda\rho}-\frac{1}{d-2}\big(2\eta^{\mu\nu}\partial_\sigma\xi{}_2^{\sigma\lambda\rho}
+\eta^{\mu\lambda}\partial_\sigma\xi{}_2^{\nu\sigma\rho}\\[2mm]
\displaystyle +\,\eta^{\nu\lambda}\partial_\sigma\xi{}_2^{\mu\sigma\rho}+\eta^{\mu\rho}\partial_\sigma\xi{}_2^{\nu\lambda\sigma}+\eta^{\nu\rho}\partial_\sigma\xi{}_2^{\mu\lambda\sigma}\big)\,,
\end{array}
\end{equation}
where the gauge parameters $\widetilde{\xi}_1$ and $\xi_2$ are arbitrary traceless hook tensors and totally antisymmetric tensors, respectively. The dual equations (\ref{dual-spin-2m}) enjoy gauge identities (\ref{GI-Dual2}),
\begin{equation}
\begin{array}{c}
\displaystyle \frac{1}{d-1}\eta^{\mu\nu}\overline{E}_{\mu\nu}-\frac{d-2}{d-1}m^{-2}\partial^{\mu}\partial^{\nu}\overline{E}_{\mu\nu}=0\,;\\[2mm]
\displaystyle -\,\frac{d-2}{d}m^{-2}\partial_\alpha\partial^\mu\partial^\nu \overline{E}_{\mu\nu}-\partial^\nu \overline{E}_{\mu\nu}+\frac{1}{d}\eta^{\mu\nu}\partial_\alpha \overline{E}_{\mu\nu}=0\,.
\end{array}
\end{equation}

As in the case of massive spin one, using the involutive closure (\ref{Inv1-spin-2m}), one can construct the``parent action" (\ref{SSt}). This  Stueckelberg action for massive spin two take the form
\begin{equation}\label{SSt-spin-2m}
\begin{array}{c}
\displaystyle \mathcal{S}[\phi,\sigma,\omega]=\int d^4x\, \mathfrak{L}\,, \quad \mathfrak{L}=\frac{1}{2}\big(h^{\mu\nu}+\frac{1}{d-1}\eta^{\mu\nu}\sigma+\frac{d-2}{d-1}m^{-2}\partial^{\mu}\partial^{\nu}\sigma+\frac{1}{2}(\partial^\nu\sigma^\mu+\partial^\mu\sigma^\nu)
\\[2mm]
\displaystyle -\,\frac{1}{d}\eta^{\mu\nu}\partial_{\alpha'}\sigma^{\alpha'}-\frac{d-2}{d}m^{-2}\partial^\mu\partial^\nu\partial_{\alpha'}\sigma^{\alpha'}+\partial_{\lambda'}\partial_{\rho'}\widetilde{\omega}{}^{\mu\nu\lambda'\rho'}
\big)\big[(\square+m^2)(\eta_{\mu\nu}h-h_{\mu\nu}\partial_\lambda\partial_\rho
\widetilde{\omega}^{\mu\nu\lambda\rho})
\\[2mm]
\displaystyle -\,\partial_\mu\partial_\nu h-\eta_{\mu\nu}\partial_\lambda\partial_\rho h^{\lambda\rho}+\partial^\lambda(\partial_\mu h_{\nu\lambda}+\partial_\nu h_{\mu\lambda})+m^2\big(\eta_{\lambda\rho}(\sigma-\frac{1}{d}\partial_\alpha\sigma^\alpha)-\frac{1}{2}(\partial_\rho\sigma_\lambda+\partial_\lambda\sigma_\rho)
\big)\big]\,,
\end{array}
\end{equation}
where $\sigma$, $\sigma^\mu$ and $\widetilde{\omega}^{\mu\nu\lambda\rho}$ (\ref{omega-spin-2m}) play the role of Stueckelberg fields. The action is invariant under gauge symmetry transformations (cf. (\ref{StGS}), (\ref{StGS-reduce-1})--(\ref{StGS-reduce-k}))
\begin{equation}
\begin{array}{c}
\displaystyle \delta^{(0)}h^{\mu\nu}=-\,\frac{1}{d-1}\eta^{\mu\nu}\mathcal{E}+\frac{d-2}{d-1}m^{-2}\partial^{\mu}\partial^{\nu}\mathcal{E}-\frac{d-2}{d}m^{-2}\partial^\mu\partial^\nu\partial_\alpha\mathcal{E}^\alpha-\frac{1}{2}(\partial^\nu\mathcal{E}^\mu+\partial^\mu\mathcal{E}^\nu)\\[2mm]
\displaystyle +\,\frac{1}{d}\eta^{\mu\nu}\partial_\alpha\mathcal{E}^\alpha-\partial_\lambda\partial_\rho\widetilde{\mathcal{E}}{}^{\mu\nu\lambda\rho}\,, \quad \delta^{(0)}\sigma=\mathcal{E}\,, \quad \delta^{(0)}\sigma^\alpha=\mathcal{E}^\alpha\,,
\quad \delta^{(0)}\widetilde{\omega}{}^{\mu\nu\lambda\rho}=\widetilde{\mathcal{E}}{}^{\mu\nu\lambda\rho}-\partial^\lambda\partial_\sigma\widetilde{\xi}{}^{\mu\nu\rho\sigma}
\\[2mm]
\displaystyle -\,\partial^\rho\partial_\sigma\widetilde{\xi}{}^{\mu\nu\lambda\sigma}+\partial^\mu\partial_\sigma\widetilde{\xi}{}^{\nu\lambda\rho\sigma}+\partial^\mu\partial_\sigma\widetilde{\xi}{}^{\nu\rho\lambda\sigma}+\partial^\nu\partial_\sigma\widetilde{\xi}{}^{\mu\lambda\rho\sigma}+\partial^\nu\partial_\sigma\widetilde{\xi}{}^{\mu\rho\lambda\sigma}\,;
\end{array}
\end{equation}
\begin{equation}
\begin{array}{c}
\displaystyle \delta^{(1)}\widetilde{\mathcal{E}}{}^{\mu\nu\lambda\rho}=\partial^\lambda\partial_\sigma\widetilde{\mathcal{E}}{}_1^{\mu\nu\rho\sigma}+\partial^\rho\partial_\sigma\widetilde{\mathcal{E}}{}_1^{\mu\nu\lambda\sigma}-\partial^\mu\partial_\sigma\widetilde{\mathcal{E}}{}_1^{\nu\lambda\rho\sigma}-\partial^\mu\partial_\sigma\widetilde{\mathcal{E}}{}_1^{\nu\rho\lambda\sigma}-\partial^\nu\partial_\sigma\widetilde{\mathcal{E}}{}_1^{\mu\lambda\rho\sigma}-\partial^\nu\partial_\sigma\widetilde{\mathcal{E}}{}_1^{\mu\rho\lambda\sigma}\,,
\\[2mm]
\displaystyle \delta^{(1)}\mathcal{E}=0\,, \quad \delta^{(1)}\mathcal{E}^\alpha=0\,, \quad
\delta^{(1)}\widetilde{\xi}{}^{\mu\nu\lambda\rho}=\widetilde{\mathcal{E}}_1^{\mu\nu\lambda\rho}-\partial^\mu\xi{}^{\nu\lambda\rho}-\partial^\nu\xi{}^{\mu\lambda\rho}
\\[2mm]
\displaystyle +\,\frac{1}{d-2}\big(2\eta^{\mu\nu}\partial_\sigma
\xi{}^{\sigma\lambda\rho}+\eta^{\mu\lambda}\partial_\sigma
\xi{}^{\nu\sigma\rho}+\eta^{\nu\lambda}\partial_\sigma\xi{}^{\mu\sigma\rho}+\eta^{\mu\rho}\partial_\sigma\xi{}^{\nu\lambda\sigma}+\eta^{\nu\rho}\partial_\sigma\xi{}^{\mu\lambda\sigma}\big)\,;
\end{array}
\end{equation}
\begin{equation}
\begin{array}{c}
\displaystyle \delta^{(2)}\mathcal{E}{}_1^{\mu\nu\lambda\rho}=\partial^\mu\mathcal{E}{}^{\nu\lambda\rho}+\partial^\nu\mathcal{E}{}^{\mu\lambda\rho}-\frac{1}{d-2}\big(2\eta^{\mu\nu}\partial_\sigma\mathcal{E}{}^{\sigma\lambda\rho}+\eta^{\mu\lambda}\partial_\sigma\mathcal{E}{}^{\nu\sigma\rho}
\\[2mm]
\displaystyle +\,\eta^{\nu\lambda}\partial_\sigma\mathcal{E}{}^{\mu\sigma\rho}+\eta^{\mu\rho}\partial_\sigma\mathcal{E}{}^{\nu\lambda\sigma}+\eta^{\nu\rho}\partial_\sigma\mathcal{E}{}^{\mu\lambda\sigma}\big)\,, \quad \delta^{(2)}\widetilde{\xi}{}^{\mu\nu\lambda}=\mathcal{E}{}^{\mu\nu\lambda}\,.
\end{array}
\end{equation}

Gauge-fixing condition (cf. (\ref{gauge1}))
\begin{equation}\label{gauge1-spin-2m}
\displaystyle \sigma=0\,, \quad \sigma^\mu=0\,, \quad \widetilde{\omega}^{\mu\nu\lambda\delta}=0\,,
\end{equation}
kills all the Stueckelberg fields and reproduce original theory (\ref{EoMs-spin-2m}).  Another admissible gauge (cf. (\ref{gauge2})) removes $\sigma$, $\sigma^\mu$ and the original fields $h^{\mu\nu}$,
\begin{equation}\label{gauge2-spin-2m-1}
\displaystyle h^{\mu\nu}=0\,, \quad \sigma=0\,, \quad \sigma^\mu=0\,.
\end{equation}
It leads to dual formulation (\ref{dual-spin-2m}) for massive spin two fields in terms of $\widetilde{\omega}{}^{\mu\nu\lambda\rho}$. The residual gauge symmetry for $\widetilde{\omega}{}^{\mu\nu\lambda\rho}$ can be fixed by gauge condition
\begin{equation}\label{gauge2-spin-2m-2}
\begin{array}{c}
\displaystyle \frac{2}{3}
\Big(\partial_\alpha\big(2(\partial^{[\rho}\widetilde{\omega}{}^{\lambda]\alpha\mu\nu}-\partial^{[\rho}\widetilde{\omega}{}^{\lambda](\mu\nu)\alpha})+\partial^\rho(\widetilde{\omega}{}^{\mu\nu\lambda\alpha}-\widetilde{\omega}{}^{\alpha(\mu\nu)\lambda})-\partial^\lambda(\widetilde{\omega}{}^{\mu\nu\rho\alpha}-\widetilde{\omega}{}^{\alpha(\mu\nu)\rho})\big)
\\[2mm]
\displaystyle +\,\frac{1}{d-1}\eta_{\alpha\beta}\partial_\gamma
\big[\eta^{\mu\nu}\big(2(\partial^{[\rho}\widetilde{\omega}{}^{\lambda]\alpha\beta\gamma}-\partial^{[\rho}\widetilde{\omega}{}^{\delta]\gamma\alpha\beta})+\partial^\rho\widetilde{\omega}{}^{\gamma\alpha\beta\lambda}-\partial^\lambda\widetilde{\omega}{}^{\gamma\alpha\beta\rho}\big)
\\[2mm]
\displaystyle
-\eta^{\nu[\lambda}\partial^{\rho]}\big(\widetilde{\omega}{}^{\mu\alpha\beta\gamma}+\widetilde{\omega}{}^{\gamma\alpha\beta\mu}-\widetilde{\omega}{}^{\mu\gamma\alpha\beta}\big)-\eta^{\mu[\lambda}\partial^{\rho]}\big(\widetilde{\omega}{}^{\nu\alpha\beta\gamma}+\widetilde{\omega}{}^{\gamma\alpha\beta\nu}-\widetilde{\omega}{}^{\nu\gamma\alpha\beta}\big)\big]\Big)=0\,.
\end{array}
\end{equation}

So, the massive spin two can be described in terms of traceless window tensor $\widetilde{\omega}{}^{\mu\nu\lambda\rho}$, being the potential for $h^{\mu\nu}$, subject to the fourth order non-Lagrangian equations (\ref{dual-spin-2m}). The dual formulations (\ref{EoMs-spin-2m}) and (\ref{dual-spin-2m}) can be derived from a single Stueckelberg action (\ref{SSt-spin-2m}) by the appropriate choice of gauge-fixing conditions: (\ref{gauge1-spin-2m}) or (\ref{gauge2-spin-2m-1})--(\ref{gauge2-spin-2m-2}), respectively.

\subsection{Massless spin $2$.}

Consider massless spin two field in $d$-dimensional Minkowski space,
\begin{equation}\label{S-spin-2}
\displaystyle S=\int d^dx\,\mathcal{L}\,, \quad \mathcal{L}=\frac{1}{4}\Big(h\big(\square h-\partial_\mu\partial_\nu h^{\mu\nu}\big)+h^{\mu\nu}\big(\partial^\lambda(\partial_\nu h_{\mu\lambda}+\partial_\mu h_{\nu\lambda})-\square h_{\mu\nu}-\partial_\mu\partial_\nu h\big)\Big)\,.
\end{equation}
Here, $h^{\mu\nu}$ is a symmetric second rank tensor, $h=\eta_{\mu\nu}h^{\mu\nu}$, and operator $\hat{M}$ (cf. (\ref{M})) reads
\begin{equation}
\begin{array}{c}
\displaystyle \hat{M}_{\mu\nu\lambda\rho}=\frac{1}{2}\Big(\big(\eta_{\mu\nu}\eta_{\lambda\rho}-\frac{1}{2}(\eta_{\mu\lambda}\eta_{\nu\rho}+\eta_{\nu\lambda}\eta_{\mu\rho})\big)\square
-\eta_{\lambda\rho}\partial_\mu\partial_\nu
-\eta_{\mu\nu}\partial_\lambda\partial_\rho
\\[2mm]
\displaystyle +\,\frac{1}{2}\big(\eta_{\mu\lambda}\partial_\nu\partial_\rho+\eta_{\nu\lambda}\partial_\mu\partial_\rho+\eta_{\mu\rho}\partial_\nu\partial_\lambda+\eta_{\nu\rho}\partial_\mu\partial_\lambda\big)\Big)\,.
\end{array}
\end{equation}
Corresponding EoMs are linearised Einstein's equations,
\begin{equation}\label{EoMs-spin-2}
\displaystyle E_{\mu\nu}\equiv\frac{\delta S}{\delta h^{\mu\nu}}=\frac{1}{2}\big(\eta_{\mu\nu}(\square h-\partial_\lambda\partial_\rho h^{\lambda\rho})+\partial^\lambda(\partial_\nu h_{\mu\lambda}+\partial_\mu h_{\nu\lambda})-\square h_{\mu\nu}-\partial_\mu\partial_\nu h\big)=0\,.
\end{equation}
The EoMs are invariant under gauge symmetry transformations,
\begin{equation}
\displaystyle \delta_\epsilon h^{\mu\nu}=\partial^\mu\epsilon^\nu+\partial^\nu\epsilon^\mu\,,
\end{equation}
i.e. $\hat{R}^{\mu\nu}{}_\alpha=(\delta^\mu_\alpha
\partial^\nu+\delta^\nu_\alpha\partial^\mu)$, cf. (\ref{gtr-R}).

Linearised Einstein's equations (\ref{EoMs-spin-2}) admit the consequence
\begin{equation}\label{tau-spin-2}
\displaystyle \tau\equiv\eta^{\mu\nu}E_{\mu\nu}=\frac{d-2}{2}\big(\square h-\partial_\mu\partial_\nu h^{\mu\nu}\big)\approx0\,,
\end{equation}
being linearised Nordstr\"om equation. Here, $\hat{\Gamma}^\dagger{}^{\mu\nu}=\eta^{\mu\nu}$, cf. (\ref{diffcons}).
The Nordstr\"om equation is a topological theory \cite{Abakumova:2022eoo}, so we have the topological subsystem among Einstein's equations.

If we add the Nordstr\"om equation to Einstein's system, the system will remain involutive (\ref{Inv}). In constructing the dual theory for linearised Einstein's gravity we proceed from the involutive closure
\begin{equation}
\displaystyle E_{\mu\nu}=0\,, \quad \tau=0\,,
\end{equation}
where $E_{\mu\nu}$, $\tau$ are defined by (\ref{EoMs-spin-2}) and (\ref{tau-spin-2}), respectively.
By construction, there exist gauge identities
\begin{equation}
\displaystyle \eta^{\mu\nu}{}E_{\mu\nu}-\tau \equiv 0\,.
\end{equation}
The consequence $\tau$ (\ref{tau-spin-2}), being considered on its own, admits gauge symmetry transformation \cite{Abakumova:2022eoo},
\begin{equation}\label{gtr-spin-2}
\delta^{(0)}_\zeta h^{\mu\nu}=\partial_\lambda\zeta^{\mu\nu\lambda}-\frac{1}{d-1}\eta_{\alpha\beta}\big(\eta^{\mu\nu}\partial_\lambda\zeta^{\alpha\beta\lambda}+\partial^\nu\zeta^{\alpha\beta\mu}+\partial^\mu\zeta^{\alpha\beta\nu}\big)\,,
\end{equation}
where $\zeta^{\mu\nu\lambda}$ is a tensor with hook symmetry type,
\begin{equation}
\displaystyle
\begin{ytableau}
\phantom{1} & \phantom{1} \\ \phantom{1} \
\end{ytableau}\,:
\quad \zeta^{(\mu\nu)\lambda}=\zeta^{\mu\nu\lambda}\,,
\quad \zeta^{(\mu\nu\lambda)}=0\,.
\end{equation}
The generator of gauge transformation (\ref{gtr-spin-2}) reads
\begin{equation}\label{rho-0-spin-2}
\begin{array}{c}
\displaystyle \hat{\rho}^{\mu\nu}{}_{\alpha\beta\gamma}=
\frac{1}{3}\Big(
2\delta^\mu_{(\alpha}\delta^\nu_{\beta)}\partial_\gamma-
\delta^\mu_{(\beta}\delta^\nu_{\gamma)}\partial_\alpha-
\delta^\mu_{(\gamma}\delta^\nu_{\alpha)}\partial_\beta
-\frac{1}{d-1}\big[\eta^{\mu\nu}\big(2\eta_{\alpha\beta}\partial_\gamma-\eta_{\beta\gamma}\partial_\alpha-\eta_{\gamma\alpha}\partial_\beta\big)\\[2mm]
\displaystyle +\,\big(2\eta_{\alpha\beta}\delta^\nu_\gamma-
\eta_{\beta\gamma}\delta^\nu_\alpha-\eta_{\gamma\alpha}\delta^\nu_\beta\big)\partial^\mu+\big(2\eta_{\alpha\beta}\delta^\mu_\gamma-\eta_{\beta\gamma}\delta^\mu_\alpha
-\eta_{\gamma\alpha}\delta^\mu_\beta\big)\partial^\nu
\big]\Big)\,,
\end{array}
\end{equation}
cf. (\ref{gtr-0}). Gauge transformation (\ref{gtr-spin-2}) is a general solution to the linearised Nordstr\"om equation (\ref{tau-spin-2}), being the trace of Einstein's equations (\ref{EoMs-spin-2}). The transformations (\ref{gtr-spin-2}) are reducible,
\begin{equation}\label{gtr-1-spin-2}
\displaystyle \delta^{(1)}_{\zeta_1}\zeta^{\mu\nu\lambda}=\partial_\rho\big(\zeta_1^{\mu\nu\lambda\rho}-\frac{1}{3}\frac{1}{d-1}\eta_{\alpha\beta}(2\eta^{\mu\nu}\zeta_1^{\alpha\beta\lambda\rho}-\eta^{\nu\lambda}\zeta_1^{\alpha\beta\mu\rho}-\eta^{\lambda\mu}\zeta_1^{\alpha\beta\nu\rho})\big)\,,
\end{equation}
where $\zeta_1^{\mu\nu\lambda\rho}$ is also a tensor with the hook symmetry,
\begin{equation}
\displaystyle
\begin{ytableau}
\phantom{1} & \phantom{1} \\ \phantom{1} \\ \phantom{1} \
\end{ytableau}\,:
\quad \zeta_1^{(\mu\nu)\lambda\rho}=\zeta_1^{\mu\nu\lambda\rho}\,,
\quad \zeta_1^{\mu\nu[\lambda\rho]}=\zeta_1^{\mu\nu\lambda\rho}\,,
\quad \zeta_1^{(\mu\nu\lambda)\rho}=0\,.
\end{equation}
The generator of the symmetry for symmetry reads
\begin{equation}\label{rho-1-spin-2}
\begin{array}{c}
\displaystyle \hat{\rho}^{\mu\nu\lambda}{}_{\alpha\beta\gamma\delta}=\frac{1}{3}\Big(\delta^\mu_\alpha
\delta^\nu_\beta\delta^\lambda_{[\gamma}\partial_{\delta]}
+\delta^\mu_\beta\delta^\nu_\alpha
\delta^\lambda_{[\gamma}\partial_{\delta]}
-\frac{1}{2}\delta^\mu_\beta\delta^\lambda_\alpha
\delta^\nu_{[\gamma}\partial_{\delta]}
-\frac{1}{2}\delta^\nu_\beta\delta^\lambda_\alpha
\delta^\mu_{[\gamma}\partial_{\delta]}
-\frac{1}{2}\delta^\nu_\alpha\delta^\lambda_\beta
\delta^\mu_{[\gamma}\partial_{\delta]}
\\[2mm]
\displaystyle -\,\frac{1}{2}\delta^\mu_\alpha\delta^\lambda_\beta
\delta^\nu_{[\gamma}\partial_{\delta]}-\frac{1}{3}\frac{1}{d-1}\big[2\eta^{\mu\nu}\big(2\eta_{\alpha\beta}\delta^\lambda_{[\gamma}
\partial_{\delta]}-\delta^\lambda_\alpha\eta_{\beta[\gamma}
\partial_{\delta]}-\delta^\lambda_\beta\eta_{\alpha[\gamma}\partial_{\delta]}\big)
\\[2mm]
\displaystyle -\,\eta^{\nu\lambda}\big(2\eta_{\alpha\beta}\delta^\mu_{[\gamma}
\partial_{\delta]}-\delta^\mu_\alpha\eta_{\beta[\gamma}\partial_{\delta]}-\delta^\mu_\beta\eta_{\alpha[\gamma}\partial_{\delta]}\big)-\eta^{\mu\lambda}\big(2\eta_{\alpha\beta}\delta^\nu_{[\gamma}
\partial_{\delta]}-\delta^\nu_\alpha\eta_{\beta[\gamma}\partial_{\delta]}-\delta^\nu_\beta\eta_{\alpha[\gamma}\partial_{\delta]}\big)\big]\Big)\,,
\end{array}
\end{equation}
cf. (\ref{gtr-1}). Transformations (\ref{gtr-1-spin-2}) admit further reducibility,
\begin{equation}\label{gtr-k-spin-2}
\displaystyle \displaystyle \delta^{(k)}_{\zeta_k}\zeta_{k-1}^{\mu\nu\lambda\rho_1\ldots\rho_{k-1}}=\partial_{\rho_k}\zeta_k^{\mu\nu\lambda\rho_1\ldots\rho_k}\,, \quad k=2,\ldots,d-2\,,
\end{equation}
cf. (\ref{gtr-2})--(\ref{gtr-k}), where
\begin{equation}
\begin{array}{c}
\displaystyle \hat{\rho}^{\mu\nu\lambda\rho_1\ldots\rho_{k-1}}{}_{\alpha\beta\gamma\delta_1\ldots\delta_k}=\frac{1}{3}\Big(\delta^\mu_\alpha
\delta^\nu_\beta\delta^\lambda_{[\gamma}
+\delta^\mu_\beta\delta^\nu_\alpha
\delta^\lambda_{[\gamma}
-\frac{1}{2}\delta^\mu_\beta\delta^\lambda_\alpha
\delta^\nu_{[\gamma}
-\frac{1}{2}\delta^\nu_\beta\delta^\lambda_\alpha
\delta^\mu_{[\gamma}
-\frac{1}{2}\delta^\nu_\alpha\delta^\lambda_\beta
\delta^\mu_{[\gamma}
\\[2mm]
\displaystyle -\,\frac{1}{2}
\delta^\mu_\alpha\delta^\lambda_\beta
\delta^\nu_{[\gamma}-\frac{1}{3}\frac{1}{d-1}\big[2\eta^{\mu\nu}\big(2\eta_{\alpha\beta}\delta^\lambda_{[\gamma}
-\delta^\lambda_\alpha\eta_{\beta[\gamma}
-\delta^\lambda_\beta\eta_{\alpha[\gamma}\big)
-\eta^{\nu\lambda}\big(2\eta_{\alpha\beta}\delta^\mu_{[\gamma}
-\delta^\mu_\alpha\eta_{\beta[\gamma}
-\delta^\mu_\beta\eta_{\alpha[\gamma}\big)
\\[2mm]
\displaystyle -\,\eta^{\mu\lambda}\big(2\eta_{\alpha\beta}\delta^\nu_{[\gamma}
-\delta^\nu_\alpha\eta_{\beta[\gamma}-\delta^\nu_\beta\eta_{\alpha[\gamma}\big)\,
\big]\Big)\delta^{\rho_1}_{\delta_1}\ldots\delta^{\rho_{k-1}}_{\delta_{k-1}}\partial_{\delta_k]}\,, \quad k=2,\ldots, d-2\,.
\end{array}
\end{equation}
The gauge parameters are hook symmetry type tensors,
\begin{equation}
\begin{array}{c}
\displaystyle \zeta_k^{(\mu\nu)\lambda\rho_1\ldots\rho_k}=\zeta_k^{\mu\nu\lambda\rho_1\ldots\rho_k}: \quad \zeta_k^{\mu\nu[\lambda\rho_1\ldots\rho_k]}=\zeta_k^{\mu\nu\lambda\rho_1\ldots\rho_k}\,, \quad
\zeta_k^{(\mu\nu\lambda)\rho_1\ldots\rho_k}=0\,, \quad
k=2,\ldots,d-2\,.
\end{array}
\end{equation}
The complete sequence of gauge symmetry transformations for linearised Nordstr\"om equation was first found in \cite{Abakumova:2022eoo}. Counting the DoF number, one can verify that linearised Nordstr\"om equation (\ref{tau-spin-2}) with reducible gauge symmetry transformations (\ref{gtr-spin-2}), (\ref{gtr-1-spin-2}), (\ref{gtr-k-spin-2}) is a topological theory indeed.

The higher order consequences, being variations of the original action (\ref{S-spin-2}) with respect to the gauge transformations (\ref{gtr-spin-2}) of topological equation (\ref{tau-spin-2}), read
\begin{equation}\label{T-spin2}
\begin{array}{c}
\displaystyle \mathcal{T}_{\alpha\beta\gamma}\equiv
\frac{1}{6}
\big(2\partial_\alpha\partial_\beta\partial^\lambda h_{\gamma\lambda}-\partial_\beta\partial_\gamma
\partial^\lambda h_{\alpha\lambda}
-\partial_\gamma\partial_\alpha\partial^\lambda h_{\beta\lambda}+2\square\partial_\gamma h_{\alpha\beta}
-\square\partial_\alpha h_{\beta\gamma}
-\square\partial_\beta h_{\gamma\alpha}\big)
\\[2mm]
\displaystyle -\,\frac{1}{3}(2\eta_{\alpha\beta}\partial_\gamma-\eta_{\beta\gamma}\partial_{\alpha}
-\eta_{\gamma\alpha}\partial_\beta)(\square h-\partial_\lambda\partial_\rho h^{\lambda\rho})\approx 0\,,
\end{array}
\end{equation}
cf. (\ref{TA}), where $\hat{\rho}^\dagger_{\alpha\beta\gamma}{}^{\mu\nu}$ is a conjugate to $\hat{\rho}_{\alpha\beta\gamma}{}^{\mu\nu}$ (\ref{rho-0-spin-2}). By construction, there exist gauge identities
\begin{equation}
\begin{array}{c}
\displaystyle \mathcal{T}_{\alpha\beta\gamma}+\frac{1}{3}\big(2\partial_\gamma E_{\alpha\beta}-\partial_\alpha E_{\beta\gamma}-\partial_\beta E_{\gamma\alpha}
-\frac{1}{d-1}[(2\eta_{\alpha\beta}\partial_\gamma-\eta_{\beta\gamma}\partial_\alpha-\eta_{\gamma\alpha}\partial_\beta)\eta^{\mu\nu}E_{\mu\nu}
\\[2mm]
\displaystyle +\,4\eta_{\alpha\beta}\partial^\lambda E_{\gamma\lambda}-2\eta_{\beta\gamma}\partial^\lambda E_{\alpha\lambda}-2\eta_{\gamma\alpha}\partial^\lambda E_{\beta\lambda}]\big)\equiv 0\,,
\end{array}
\end{equation}
cf. (\ref{GI-TA}). The identities (\ref{TA-1-2}) between consequences $\mathcal{T}_{\alpha\beta\gamma}$ read
\begin{equation}
\begin{array}{c}
\displaystyle -\,\frac{1}{3}\Big(\partial_\delta\mathcal{T}_{\alpha\beta\gamma}-\partial_\gamma\mathcal{T}_{\alpha\beta\delta}-\frac{1}{4}\partial_\delta(\mathcal{T}_{\beta\gamma\alpha}+\mathcal{T}_{\gamma\beta\alpha}+\mathcal{T}_{\gamma\alpha\beta}+\mathcal{T}_{\alpha\gamma\beta})
+\frac{1}{4}\partial_\gamma(\mathcal{T}_{\beta\delta\alpha}+\mathcal{T}_{\delta\beta\alpha}
+\mathcal{T}_{\delta\alpha\beta}
\\[2mm]
\displaystyle +\,\partial_\gamma\mathcal{T}_{\alpha\delta\beta})
-\frac{1}{3}\frac{1}{d-1}\big[2\eta^{\mu\nu}\big(\eta_{\alpha\beta}(\partial_\delta\mathcal{T}_{\mu\nu\gamma}-\partial_\gamma
\mathcal{T}_{\mu\nu\delta})-\eta_{\beta[\gamma}
\partial_{\delta]}\mathcal{T}_{\mu\nu\alpha}
-\eta_{\alpha[\gamma}\partial_{\delta]}\mathcal{T}_{\mu\nu\beta}\big)
\\[2mm]
\displaystyle -\,\eta^{\nu\lambda}\big(\eta_{\alpha\beta}(\partial_{\delta}\mathcal{T}_{\gamma\nu\lambda}-\partial_\gamma\mathcal{T}_{\delta\nu\lambda})-\eta_{\beta[\gamma}\partial_{\delta]}\mathcal{T}_{\alpha\nu\lambda}-\eta_{\alpha[\gamma}\partial_{\delta]}\mathcal{T}_{\beta\nu\lambda}\big)
\\[2mm]
\displaystyle -\,\eta^{\mu\lambda}\big(\eta_{\alpha\beta}(\partial_\delta\mathcal{T}_{\mu\gamma\lambda}-\partial_\gamma\mathcal{T}_{\mu\delta\lambda})-\eta_{\beta[\gamma}\partial_{\delta]}\mathcal{T}_{\mu\alpha\lambda}
-\eta_{\alpha[\gamma}\partial_{\delta]}\mathcal{T}_{\mu\beta\lambda}\big)\big]\Big)\equiv0\,,
\end{array}
\end{equation}
as
\begin{equation}
\displaystyle \hat{\rho}^\dagger_{\alpha\beta\gamma\delta}{}^{\mu\nu\lambda}\hat{\rho}^\dagger_{\mu\nu\lambda}{}^{\alpha'\beta'}\equiv0\,,
\end{equation}
where $\hat{\rho}^\dagger_{\alpha\beta\gamma\delta}{}^{\mu\nu\lambda}$ is a conjugate to $\hat{\rho}^\dagger{}^{\mu\nu\lambda}{}_{\alpha\beta\gamma\delta}$ (\ref{rho-1-spin-2}). These identities admit further reducibility,
\begin{equation}
\begin{array}{c}
\displaystyle \hat{\rho}^\dagger_{\mu\nu\lambda\rho_1\ldots\rho_k}{}^{\alpha\beta\gamma\delta_1\ldots\delta_{k-1}}\hat{\rho}^\dagger_{\alpha\beta\gamma\delta_1\ldots\delta_{k-1}}{}^{\alpha'\beta'\gamma'\delta'_1\ldots\delta'_{k-2}}\equiv0\,,\\
[2mm]
\displaystyle \hat{\rho}^\dagger_{\mu\nu\lambda\rho_1\ldots\rho_k}{}^{\alpha\beta\gamma\delta_1\ldots\delta_{k-1}}=-\,\frac{1}{3}\Big(\delta^\alpha_\mu
\delta^\beta_\nu\delta^\gamma_{[\lambda}
+\delta^\alpha_\nu\delta^\beta_\mu
\delta^\gamma_{[\lambda}
-\frac{1}{2}\delta^\alpha_\nu\delta^\gamma_\mu
\delta^\beta_{[\lambda}
-\frac{1}{2}\delta^\beta_\nu\delta^\gamma_\mu
\delta^\nu_{[\lambda}
-\frac{1}{2}\delta^\beta_\mu\delta^\gamma_\nu
\delta^\alpha_{[\lambda}
\\[2mm]
\displaystyle -\,\frac{1}{2}
\delta^\alpha_\mu\delta^\gamma_\nu
\delta^\beta_{[\lambda}-\frac{1}{3}\frac{1}{d-1}\big[2\eta^{\alpha\beta}(2\eta_{\mu\nu}\delta^\gamma_{[\lambda}
-\delta^\gamma_\mu\eta_{\nu[\lambda}
-\delta^\gamma_\nu\eta_{\mu[\lambda})
-\eta^{\beta\gamma}(2\eta_{\mu\nu}\delta^\alpha_{[\lambda}
-\delta^\alpha_\mu\eta_{\nu[\lambda}
-\delta^\alpha_\nu\eta_{\mu[\lambda})
\\[2mm]
\displaystyle -\,\eta^{\alpha\gamma}(2\eta_{\mu\nu}\delta^\beta_{[\lambda}
-\delta^\beta_\mu\eta_{\nu[\lambda}-\delta^\beta_\nu\eta_{\mu[\lambda})\,
\big]\Big)\delta^{\delta_1}_{\rho_1}\ldots\delta^{\delta_{k-1}}_{\rho_{k-1}}\partial_{\rho_k]}\,,
\end{array}
\end{equation}
where $k=2,\dots,d-2$.

The involutive closure (\ref{Inv1}) reads
\begin{equation}\label{Inv1-spin-2}
\displaystyle E_{\mu\nu}=0\,, \quad \tau=0\,, \quad \mathcal{T}_{\alpha\beta\gamma}=0\,,
\end{equation}
where $E_{\mu\nu}$, $\tau$, $\mathcal{T}_{\alpha\beta\gamma}$ are defined by (\ref{EoMs-spin-2}), (\ref{tau-spin-2}) and (\ref{T-spin2}), respectively.

The general solution for the Nordst\"om equation (\ref{tau-spin-2}) reads
\begin{equation}\label{gensol-Nord}
\displaystyle h^{\mu\nu}=\partial_\lambda\omega^{\mu\nu\lambda}-\frac{1}{d-1}\eta_{\alpha\beta}\big(\eta^{\mu\nu}\partial_\lambda\omega^{\alpha\beta\lambda}+\partial^\nu\omega^{\alpha\beta\mu}+\partial^\mu\omega^{\alpha\beta\nu}\big)\,,
\end{equation}
where $\omega^{\mu\nu\lambda}$ is a hook symmetry type tensor,
\begin{equation}\label{omega-spin-2}
\displaystyle
\begin{ytableau}
\phantom{1} & \phantom{1} \\ \phantom{1} \
\end{ytableau}\,:
\quad \omega^{(\mu\nu)\lambda}=\omega^{\mu\nu\lambda}\,,
\quad \omega^{(\mu\nu\lambda)}=0\,.
\end{equation}
The dual equations for massless spin two are constructed by the recipe (\ref{dual}), i.e. by substituting the general solution (\ref{gensol-Nord}) of Nordstr\"om equation into linearised Einstein's system (\ref{EoMs-spin-2}),
\begin{equation}\label{dual-spin-2}
\overline{E}_{\mu\nu}\equiv\frac{1}{2}
\big(\partial_\mu\partial^\lambda\partial^\rho\omega_{\nu\lambda\rho}+\partial_\nu\partial^\lambda\partial^\rho\omega_{\mu\lambda\rho}-\square\partial^\lambda\omega_{\mu\nu\lambda}\big)+\eta_{\alpha\beta}\big(\eta_{\mu\nu}\square-\partial_\mu\partial_\nu)\partial_\lambda\omega^{\alpha\beta\lambda}=0\,.
\end{equation}
The equations (\ref{dual-spin-2}) are invariant under reducible gauge symmetry transformations
\begin{equation}
\displaystyle \delta^{(0)}_{\xi_1}\omega^{\mu\nu\lambda}=\partial_\rho\big(\zeta_2^{\mu\nu\lambda\rho}-\frac{1}{3}\frac{1}{d-1}\eta_{\alpha\beta}(2\eta^{\mu\nu}\zeta_2^{\alpha\beta\lambda\rho}-\eta^{\nu\lambda}\zeta_2^{\alpha\beta\mu\rho}-\eta^{\lambda\mu}\zeta^{\alpha\beta\nu\rho})\big)\,;
\end{equation}
\begin{equation}
\displaystyle \delta^{(k-1)}_{\xi_k}\xi_{k-1}^{\mu\nu\lambda\rho_1\ldots\rho_{k-1}}=\partial_{\rho_k}\xi_k^{\mu\nu\lambda\rho_1\ldots\rho_{k}}\,,
\end{equation}
where $k=2,\ldots, d-2$, and the gauge parameters are arbitrary tensors with hook symmetry type,
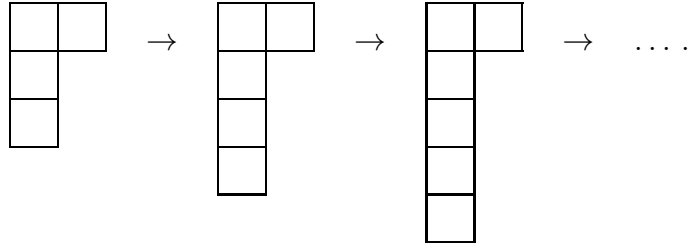
\begin{equation*}
\begin{ytableau}
\phantom{1} & \phantom{1} \\ \phantom{1} \\ \phantom{1} \
\end{ytableau}
\quad \rightarrow \quad
\begin{ytableau}
\phantom{1} & \phantom{1} \\ \phantom{1} \\ \phantom{1} \\ \phantom{1} \
\end{ytableau}
\quad \rightarrow \quad
\begin{ytableau}
\phantom{1} & \phantom{1} \\ \phantom{1} \\ \phantom{1} \\ \phantom{1} \\ \phantom{1} \
\end{ytableau}
\quad \rightarrow \quad \ldots \,.
\end{equation*}
The gauge identities (\ref{GI-Dual1})--(\ref{GI-Dual2}) for dual equations (\ref{dual-spin-2}) take the form
\begin{equation}
\displaystyle \eta^{\mu\nu}\overline{E}_{\mu\nu}\equiv0\,, \quad
-\,2\partial^\nu\overline{E}_{\mu\nu}\equiv 0\,.
\end{equation}
Given all the ingredients for the Stueckelberg action (\ref{SSt}), we can construct it for the massless spin two:
\begin{equation}\label{SSt-spin-2}
\begin{array}{c}
\displaystyle \mathcal{S}[\phi,\sigma,\omega]=\int d^dx\, \mathfrak{L}\,, \quad \mathfrak{L}=\frac{1}{4}\big(h^{\mu\nu}+\eta^{\mu\nu}\sigma+\partial_{\lambda'}\omega^{\mu\nu\lambda'}-\frac{1}{d-1}\eta_{\alpha'\beta'}(\eta^{\mu\nu}\partial_{\lambda'}\omega^{\alpha'\beta'\lambda'}
\\[2mm]
\displaystyle+\,\partial^\nu\omega^{\alpha'\beta'\mu}+\partial^\mu\omega^{\alpha'\beta'\nu})\big)\big[(\eta_{\mu\nu}\square-\partial_\mu\partial_\nu)\big(h+(d-2)\sigma+2\eta_{\alpha\beta}\partial_\lambda\omega^{\alpha\beta\lambda}\big)-\square(h_{\mu\nu}+\partial^\lambda\omega_{\mu\nu\lambda})
\\[2mm]
\displaystyle -\,\eta_{\mu\nu}\partial_\lambda\partial_\rho h^{\lambda\rho}+\partial^\lambda\big(\partial_\nu(h_{\mu\lambda}+\partial^\rho\omega_{\mu\lambda\rho})+\partial_\mu(h_{\nu\lambda}+\partial^\rho\omega_{\nu\lambda\rho})\big)\big]\,,
\end{array}
\end{equation}
where $\sigma$ and $\omega^{\mu\nu\lambda}$ (\ref{omega-spin-2}) play the role of Stueckelberg fields. It is invariant under gauge symmetry transformations (cf. (\ref{StGS}), (\ref{StGS-reduce-1})--(\ref{StGS-reduce-k}))
\begin{equation}
\begin{array}{c}
\displaystyle \delta^{(0)}h^{\mu\nu}=-\,\eta^{\mu\nu}\mathcal{E}-\partial_\lambda\mathcal{E}^{\mu\nu\lambda}+\frac{1}{d-1}\eta_{\alpha\beta}\big(\eta^{\mu\nu}\partial_\lambda\mathcal{E}^{\alpha\beta\lambda}+\partial^\nu\mathcal{E}^{\alpha\beta\mu}+\partial^\mu\mathcal{E}^{\alpha\beta\nu}\big)\,, \quad \delta^{(0)}\sigma=\mathcal{E}\,,\\[2mm]
\displaystyle \delta^{(0)}\omega^{\mu\nu\lambda}=\mathcal{E}^{\mu\nu\lambda}-\partial_\rho\big(\xi^{\mu\nu\lambda\rho}-\frac{1}{3}\frac{1}{d-1}\eta_{\alpha\beta}(2\eta^{\mu\nu}\xi^{\alpha\beta\lambda\rho}-\eta^{\nu\lambda}\xi^{\alpha\beta\mu\rho}-\eta^{\lambda\mu}\xi
^{\alpha\beta\nu\rho})\big)\,.
\end{array}
\end{equation}
These gauge transformations are reducible, the sequence of symmetry transformations of gauge parameters reads
\begin{equation}
\begin{array}{c}
\displaystyle \delta^{(1)}\mathcal{E}^{\mu\nu\lambda}=\partial_\rho\big(\mathcal{E}^{\mu\nu\lambda\rho}-\frac{1}{3}\frac{1}{d-1}\eta_{\alpha\beta}(2\eta^{\mu\nu}\mathcal{E}^{\alpha\beta\lambda\rho}-\eta^{\nu\lambda}\mathcal{E}^{\alpha\beta\mu\rho}-\eta^{\lambda\mu}\mathcal{E}
^{\alpha\beta\nu\rho})\big)\,,\\[2mm]
\displaystyle \delta^{(1)}\mathcal{E}=0\,, \quad \delta^{(1)}\xi^{\mu\nu\lambda\rho}=\mathcal{E}^{\mu\nu\lambda}-\partial_{\rho_2}\xi^{\mu\nu\lambda\rho_1\rho_2}\,;
\end{array}
\end{equation}
\begin{equation}
\displaystyle \delta^{(k)}\mathcal{E}^{\mu\nu\lambda\rho_1\ldots\rho_{k-1}}=\partial_{\rho_k}\mathcal{E}^{\mu\nu\lambda\rho_1\ldots\rho_k}\,, \quad
\delta^{(k)}\xi^{\mu\nu\lambda\rho_1\ldots\rho_k}=\mathcal{E}^{\mu\nu\lambda\rho_1\ldots\rho_k}-\partial_{\rho_{k+1}}\xi^{\mu\nu\lambda\rho_1\ldots\rho_{k+1}}\,,
\end{equation}
where $k=2,\ldots,d-3$;
\begin{equation}
\displaystyle \delta^{(d-2)}\mathcal{E}^{\mu\nu\lambda\rho_1\ldots\rho_{d-3}}=\partial_{\rho_{d-2}}\mathcal{E}^{\mu\nu\lambda\rho_1\ldots\rho_{d-2}}\,, \quad \delta^{(d-2)}\xi^{\mu\nu\lambda\rho_1\ldots\rho_{d-2}}=\mathcal{E}^{\mu\nu\lambda\rho_1\ldots\rho_{d-2}}\,.
\end{equation}
Gauge-fixing condition 
\begin{equation}\label{gauge1-spin-2}
\displaystyle \sigma=0\,, \quad \omega^{\mu\nu\lambda}=0\,,
\end{equation}
kills all the Stueckelberg fields and reproduces original theory (\ref{EoMs-spin-2}). Another admissible gauge (cf. (\ref{gauge2})) removes  $\sigma$ and the original field $h^{\mu\nu}$ and leads to dual formulation (\ref{dual-spin-2}) for massless spin two in terms of $\omega^{\mu\nu\lambda}$,
\begin{equation}\label{gauge2-spin-2-1}
\displaystyle h^{\mu\nu}=0\,, \quad \sigma=0\,.
\end{equation}
The residual gauge symmetry for $\omega^{\mu\nu\lambda}$ can be fixed by imposing condition
\begin{equation}\label{gauge2-spin-2-2}
\begin{array}{c}
\displaystyle -\,\frac{1}{3}\Big(\partial^\rho\omega^{\mu\nu\lambda}-\partial^\lambda\omega^{\mu\nu\rho}-\frac{1}{4}\partial^\rho(\omega^{\nu\lambda\mu}+\omega^{\lambda\nu\mu}+\omega^{\lambda\mu\nu}+\omega^{\mu\lambda\nu})
+\frac{1}{4}\partial^\lambda
(\omega^{\nu\rho\mu}+\omega^{\rho\nu\mu}+\omega^{\rho\mu\nu}
\\[2mm]
\displaystyle +\,\partial^\lambda\omega^{\mu\rho\nu})
-\frac{1}{3}\frac{1}{d-1}\big[2\eta_{\alpha\beta}\big(\eta^{\mu\nu}(\partial^\rho
\omega^{\alpha\beta\lambda}-\partial^\lambda
\omega^{\alpha\beta\rho})-\eta^{\nu[\lambda}
\partial^{\rho]}\omega^{\alpha\beta\mu}-\eta^{\mu[\lambda}
\partial^{\rho]}\omega^{\alpha\beta\nu}\big)
\\[2mm]
\displaystyle -\,\eta_{\beta\gamma}\big(\eta^{\mu\nu}(\partial^\rho\omega^{\lambda\beta\gamma}-\partial^\lambda\omega^{\rho\beta\gamma})-\eta^{\nu[\lambda}\partial^{\rho]}\omega^{\mu\beta\gamma}-\eta^{\mu[\lambda}\partial^{\rho]}\omega^{\nu\beta\gamma}\big)
\\[2mm]
\displaystyle -\,\eta_{\alpha\gamma}\big(\eta^{\mu\nu}(\partial^\rho\omega^{\alpha\lambda\gamma}-\partial^\lambda\omega^{\alpha\rho\gamma})-\eta^{\nu[\lambda}\partial^{\rho]}\omega^{\alpha\mu\gamma}-\eta^{\mu[\lambda}\partial^{\rho]}\omega^{\alpha\nu\gamma}\big)\big]\Big)=0\,.
\end{array}
\end{equation}

The hook tensor $\omega^{\mu\nu\lambda}$, being the ``potential" for the metric $h^{\mu\nu}$ has to obey involutive third-order equations (\ref{dual-spin-2}). The equations (\ref{dual-spin-2}) are non-Lagrangian, but they can be cast into Lagrangian framework by constructing ``parent action" (\ref{SSt-spin-2}). One can switch between dual theories (\ref{EoMs-spin-2}) and (\ref{dual-spin-2}) by imposing different gauge conditions --- either (\ref{gauge1-spin-2}), or (\ref{gauge2-spin-2-1})--(\ref{gauge2-spin-2-2}).

There exist dual description of gravity in terms of the Lanczos tensor \cite{Lanczos:1949zz}, \cite{Lanczos:1962zz} being also the third rank tensor with hook symmetry. The Lanczos tensor can be thought of as a potential for the Weyl tensor \cite{Takeno:1964}. At the linearised level the Lanczos tensor can be expressed in terms of derivatives of the metric \cite{Gopal:2021lax}. Somewhat similar scenario is demonstrated in article \cite{Toth:2021dut}, where the linearised Einstein's equations are written as the first order partial differential equations in terms of the ``Fierz tensor" also having the hook Young diagram. This tensor can be viewed as a field strength with metric as its potential. Unlike the mentioned formulations, we express $h^{\mu\nu}$ in terms of the derivatives of hook tensor, see (\ref{gensol-Nord}), so the hook tensor in our dualisation scheme serves as a potential for the metric, not vice versa.

Finalising discussion of dualisation of specific models, let us remark that ``parent actions" for dual theories  (\ref{SSt-spin-1}), (\ref{SSt-spin-2m}), (\ref{SSt-spin-2}) involve higher derivatives in all the considered cases. This is a consequence of the general structure of the ``parent action" (\ref{SSt}), where the operator $\hat{M}$ of original EoMs (\ref{EoMs}) is decorated with the differential operators $\hat{\rho}$, being the generators of gauge transformations for the topological subsystem (\ref{gtr-0}). This inevitably increases the order of the derivatives in the action. Once the higher derivatives are included, stability of dynamics can be questioned. Notice that higher derivative dynamics are not necessarily unstable \cite{Kaparulin:2014vpa} even if the canonical energy is unbounded. The matter is that dynamics can admit another bounded conserved quantity which prevents instability at classical and quantum level. The stability of higher derivative dynamics can survive upon inclusion of interaction \cite{Kaparulin:2015owa}, \cite{Abakumova:2018eck},
\cite{Kaparulin:2020rqz}.  In this article, we do not elaborate on the issue of stability, it will be addressed elsewhere. We only mention that the canonical energy is unbounded off-shell for the ``parent" Lagrangians. Since energy of the original model is bounded, the dual model does not suffer from instability, as the original energy expressed in terms of potentials $\omega$ is conserved anyway.

\section{Concluding remarks and discussion}

As we have seen above, the potentials can be systematically introduced for the fields of Lagrangian free theory provided for the involutive closure of the original EoMs includes a subsystem which is a topological theory. The field equations, being reformulated in terms of potentials, are dual to the original Lagrangian theory. These two dual theories can be absorbed by a uniform parent action involving both the original fields and their potentials. In the parent theory, one can switch between the dual formulations by choosing appropriate gauge conditions which remove either original fields, or their potentials. This method, being applied to the simplest cases of massive and massless spin two theories in Fierz-Pauli formulation leads to the dual formulations in terms of potential fields being tensors with window and hook Young diagram, respectively. No doubts, the basic assumption for this dualisation scheme --- existence of the topological subsystem in the involutive closure of field equations --- is met by many other free field theories.

Let us discuss the perspectives of the proposed dualisation scheme beyond the free level. The most obvious option is to follow the usual pattern of consistent inclusion of interactions in gauge theories \cite{Barnich:1993vg} proceeding from the parent action involving both the original fields and their potentials. The result can be different, in principle, than for the original action.

Let us mention another option for this dualisation scheme at non-linear level. Let us assume that EoMs of consistent non-linear theory admit the consequences being a topological subsystem. The example is the Nordsr\"om equation $R=\Lambda$ being a consequence of Einstein's equations. If the equation is considered on its own, irrespectively to entire Einstein's system, it is a topological theory. At linearised level, this is proven in \cite{Abakumova:2022eoo} by explicitly finding the gauge symmetry. At the non-linear level, the complete gauge symmetry of the Nordsr\"om equation is still unknown, although Koiso's theorem \cite{Koiso:1979}, \cite{Besse:1987} provides indirect evidence that the system does not have local DoF, see for a discussion  \cite{ Abakumova:2022eoo}. If the complete gauge symmetry of the Nordstr\"om equation is found, one can take a variation of Einstein-Hilbert action with respect to the gauge symmetry of the Nordstr\"om equation. This would lead us to the involutive system (\ref{Inv1-spin-2}) at non-linear level. This system can serve as a starting point for iterative inclusion of Stueckelberg fields following the general procedure of \cite{Lyakhovich:2021lzy}, \cite{Abakumova:2021evc} that would result in construction of a parent action for the non-linear theory of gravity. This parent action would involve the metric and its potential. Comparing to the linearised case, the potential is expected to be a third rank tensor with hook Young diagram. Notice that inclusion of Stueckelberg fields is an unobstructed iterative procedure \cite{Lyakhovich:2021lzy}, so the parent action for the metric and its potential should exist at non-linear level  at least as a formal power series in the potential.  By construction, the parent theory should admit a gauge-fixing condition  that removes the potential restoring gravity in terms of metric. The alternative gauge-fixing can be also admissible that fixes Einstein's metric, not necessarily flat. In this way, we can see the consistent couplings of metric potentials in the curved Riemannian background. Until now, no consistent self-interactions are known for spin two in any representation, except for the symmetric tensor. The proposed representation by hook tensor seems admitting interactions. Furthermore, the parent action can admit, in principle, more options for inclusion of consistent interactions with other fields comparing to representation in terms of original fields.

\vspace{2 mm}

\subsection*{Acknowledgments.}
We thank A.~A.~Sharapov for fruitful discussions. The part of the work concerning a general procedure of constructing dual formulations for the free field theories is supported by the Foundation for the Advancement of Theoretical Physics and Mathematics ``BASIS''. The dual formulation of various spins is a part of the project supported by the Tomsk State University Development Program (Priority--2030).

\end{document}